\documentclass[]{aa}
\usepackage{epsfig} 

\newcommand{\gr}{$\gamma$-ray \,}
\newcommand{\grs}{$\gamma$-rays \,}

\begin{document}
\thesaurus{3 (11.02.2, 13.07.2)}
\title{The time averaged TeV energy spectrum of Mkn 501
of the extraordinary 1997 outburst as measured with the
stereoscopic Cherenkov telescope system of HEGRA}

\author{F.A. Aharonian\inst{1},
A.G. Akhperjanian\inst{7},
J.A.~Barrio\inst{2,3},
K.~Bernl\"ohr\inst{1,}$^*$,
H. Bojahr\inst{6},
I. Calle\inst{3},
J.L. Contreras\inst{3},
J. Cortina\inst{3},
A. Daum\inst{1},
T. Deckers\inst{5},
S. Denninghoff\inst{2},
V. Fonseca\inst{3},
J.C. Gonzalez\inst{3},
G. Heinzelmann\inst{4},
M. Hemberger\inst{1},
G. Hermann\inst{1,}$^\dag$,
M. He{\ss}\inst{1},
A. Heusler\inst{1},
W. Hofmann\inst{1},
H. Hohl\inst{6},
D. Horns\inst{4},
A. Ibarra\inst{3},
R. Kankanyan\inst{1,7},
J. Kettler\inst{1},
C. K\"ohler\inst{1},
A. Konopelko\inst{1,}$^\S$,
H. Kornmeyer\inst{2},
M. Kestel\inst{2},
D. Kranich\inst{2},
H. Krawczynski\inst{1},
H. Lampeitl\inst{1},
A. Lindner\inst{4},
E. Lorenz\inst{2},
N. Magnussen\inst{6},
H. Meyer\inst{6},
R. Mirzoyan\inst{2},
A. Moralejo\inst{3},
L. Padilla\inst{3},
M. Panter\inst{1},
D. Petry\inst{2,}$^\ddag$,
R. Plaga\inst{2},
A. Plyasheshnikov\inst{1,}$^\S$,
J. Prahl\inst{4},
G. P\"uhlhofer\inst{1},
G. Rauterberg\inst{5},
C. Renault\inst{1,}$^\#$,
W. Rhode\inst{6},
A. R\"ohring\inst{4},
V. Sahakian\inst{7},
M. Samorski\inst{5},
D. Schmele\inst{4},
F. Schr\"oder\inst{6},
W. Stamm\inst{5},
H.J. V\"olk\inst{1},
B. Wiebel-Sooth\inst{6},
C. Wiedner\inst{1},
M. Willmer\inst{5},
W. Wittek\inst{2}}
\institute{Max Planck Institut f\"ur Kernphysik,
Postfach 103980, D-69029 Heidelberg, Germany \and
Max Planck Institut f\"ur Physik, F\"ohringer Ring
6, D-80805 M\"unchen, Germany \and
Universidad Complutense, Facultad de Ciencias
F\'{\i}sicas, Ciudad Universitaria, E-28040 Madrid, Spain \and
Universit\"at Hamburg, II. Institut f\"ur
Experimentalphysik, Luruper Chaussee 149,
D-22761 Hamburg, Germany \and
Universit\"at Kiel, Institut f\"ur Kernphysik,
Leibnitzstr. 15, D-24118 Kiel, Germany \and
Universit\"at Wuppertal, Fachbereich Physik,
Gau{\ss}str.20, D-42097 Wuppertal, Germany \and
Yerevan Physics Institute, Alikhanian Br. 2, 375036 Yerevan, Armenia\\
\hspace*{-4.04mm} $^*\,$ Now at Forschungszentrum Karlsruhe,
P.O. Box 3640, D-76021 Karlsruhe, Germany\\
\hspace*{-4.04mm} $^\dag\,$ Now at Enrico Fermi Institute,
The University of Chicago, 933 East 56th Street,
Chicago, IL 60637, U.S.A.\\
\hspace*{-4.04mm} $^\ddag\,$ Now at Universidad Aut\'{o}noma de Barcelona,
Institut de F\'{i}sica d'Altes Energies, E-08193 Bellaterra, Spain\\
\hspace*{-4.04mm} $^\S\,$ On live from  
Altai State University, Dimitrov Street 66, 656099 Barnaul, Russia\\
\hspace*{-4.04mm} $^\#\,$ Now at LPNHE, Universit\'es Paris VI-VII, 4
place Jussieu, F-75252 Paris Cedex 05, France
}
\mail{Henric Krawczynski, \\Tel.: (Germany) +6221 516 471,\\
email address: Henric.Krawczynski@mpi-hd.mpg.de}
\offprints{Henric Krawczynski}
\date{Received 16 March 1999; accepted 28 June 1999}
\authorrunning{F. Aharonian et al.}
\titlerunning {The 1997 Time Averaged TeV Spectrum of Mkn~501}
\maketitle
\begin{abstract}

During the several months of the outburst of Mkn~501 in
1997 the source has been monitored in TeV \grs with the HEGRA stereoscopic
system of imaging atmospheric Cherenkov telescopes. Quite remarkably it
turned out that the shapes of the daily \gr energy
spectra remained essentially stable throughout 
the entire state of high activity
despite dramatic flux variations during this period. 
The derivation of a long term time-averaged 
energy spectrum, based on more than 38,000 detected TeV
photons, is therefore physically meaningful. 
The unprecedented \gr statistics combined with the 20\% 
energy resolution of the instrument resulted in the 
first detection of $\gamma$-rays from an extragalactic 
source well beyond 10~TeV, and the first high accuracy measurement 
of an exponential cutoff in the energy region above 5~TeV
deeply into the exponential regime.
From 500 GeV to 24 TeV 
the differential photon spectrum is well approximated  
by a power-law with an exponential cutoff: 
${\rm d} N/{\rm d} E=N_0 \, (E/1\,{\rm TeV})^{-\alpha} \, \exp{(-E/E_0)}$, with
$N_0=(10.8 ~\pm0.2_{\rm stat}~\pm2.1_{\rm sys}) 
\cdot 10^{-11} \, \rm cm^{-2} s^{-1} TeV^{-1}$,   
$\alpha=1.92 ~\pm0.03_{\rm stat} ~\pm0.20_{\rm sys}$, and 
$E_0=(6.2 ~\pm0.4_{\rm stat} ~(-1.5 ~+2.9)_{\rm sys}) \, \rm TeV$.
We summarize the methods for
the evaluation of the energy spectrum in a broad dynamical range 
which covers almost two energy decades, and study in detail
the principal sources of systematic errors.
We also discuss several important astrophysical implications of the 
observed result concerning the production and absorption 
mechanisms of \grs in the emitting jet
and the modifications of the initial spectrum of TeV radiation due to
its interaction with the diffuse extragalactic background radiation.
\end{abstract}
\keywords{ BL Lacertae objects: individual:
Mkn 501 \-- gamma-rays: observations}

\section{Introduction}
Mkn~501, an active galactic nucleus (AGN) at a redshift $z = 0.034$, was
discovered several years ago as a faint source of TeV $\gamma$-radiation
(Quinn et al.\ \cite{quin:96} ; Bradbury et al.\ \cite{brad:97}). 
In 1997 it turned into a state of high activity, unique in both its 
strength and duration. The TeV
emission of the source from March to September 1997 was characterized by a
strongly variable flux. It was on average more than three times larger
than the flux of the Crab Nebula, the strongest known persistent TeV
source in the sky. Fortunately, the time period of the outburst coincided
with the source visibility windows of several ground-based imaging
atmospheric Cherenkov telescopes (IACTs) designed for the
detection of very high energy (VHE) cosmic $\gamma$-rays. 
Thus almost continuous monitoring of Mkn~501 in TeV \grs with
several IACTs (CAT, HEGRA, TACTIC, Telescope Array, Whipple)
located in the Northern Hemisphere was possible (e.g. Protheroe et
al.\ \cite{prot:98}).

The observations of Mkn~501 by the HEGRA stereoscopic IACT system during
this long outburst made a detailed study of the temporal and spectral
characteristics of the source possible, based on an unprecedented
statistics of more than 38,000 TeV photons 
(Aharonian et al.\ \cite{ahar:99a};
hereafter Paper~1).  The ``background-free'' detection of \grs, with an
average rate of several hundred \grs per hour (against $\approx 20$
background events caused by charged cosmic rays), allowed us to determine
statistically significant signals for $\leq 5$ minute intervals during
much of the 110~h observation time, spread over 6 months. 
Moreover, it was possible to monitor the energy spectrum of the source
on a daily basis.
Within the errors the energy spectrum
maintained a constant form over the range from 1 TeV to 10 TeV. 
This was the case even though the flux
varied strongly on time scales $\leq 1$ day. We believe that this is an
important result, and it was to some extent unexpected.

The diurnal spectra exhibit a power-law shape at low energies (between 1 TeV
and several TeV), with a gradual steepening towards higher energies (Paper
1). Such a spectral form could not be unequivocally ensured in the first
analysis which was performed during the period of activity of the source, since the
systematic errors of the recently commissioned stereoscopic system of 
HEGRA were not well studied at this time.
As a consequence, the energy spectrum could not be determined 
more precisely than implied by a power law fit (Aharonian et al.\ \cite{ahar:97a}), 
even though the tendency for a gradual steepening of the observed 
spectra was noticed (Aharonian et al.\ \cite{ahar:97b}). 
The results of Paper 1 and the new results in a broader 
energy interval presented below are based on detailed  systematic
studies (see Paper 1 and Konopelko et al.\ \cite{kon:99}), and extend and supersede 
these previous  results. This allowed 
us to come to the definite conclusion that the spectrum 
determined in the energy region from 1 to 10 TeV steepens 
significantly (Paper 1).
A similar tendency has been found also by the 
Whipple (Samuelson et al.\ \cite{samuel:98}),
Telescope Array (Kajino et al.\ 1999, private communication), and
CAT (Djannati-Atai et al.\ \cite{CATcorrelation}) groups.  
Independent spectral measurements
by the HEGRA telescopes CT1 and CT2 will be published elsewhere.  

Apart from its astrophysical significance, the constancy of the spectral
shape has the important practical consequence that it allows to measure 
the spectrum with small statistical errors also 
in the energy regions below 1 TeV and above 10 TeV. 
Indeed, the low photon statistics of the detector in
both "extreme" energy bands (towards low energies basically due to the
decrease of the detector's collection area; towards high energies due to
the steep photon spectrum) can be drastically increased by using the data
accumulated over the whole period of observations.

In Sect.\ 2 we describe the HEGRA stereoscopic system and the specific form
of the data analysis, based on Monte Carlo simulations of both, the air showers
and the detection system. The data sample is the same as in Paper 1 and is 
described in Sect.\ 3. A detailed study of the systematic errors in the
spectrum derivation is contained in Sect. 4; most of this methodology
was actually developed in the context of the Mkn 501 data analysis.
We believe that this is the first study of this kind.
The experimental results are then presented in Sect.\ 5, whereas Sect.\ 6 
attempts a first discussion. This discussion concentrates on the $\gamma$-ray
results of Sect.\ 5 and what one can learn from them alone.
A multiwavelength analysis is clearly outside the scope of this paper. 
Our conclusions are
contained in Sect.\ 7.

Readers only interested in the astrophysical results, 
should skip Sect.\ 2-4 and proceed to Sect.\ 5.
\section{The HEGRA system of imaging atmospheric Cherenkov telescopes}

\subsection{The HEGRA Cherenkov telescope system}

The VHE $\gamma$-ray observatory of the HEGRA collaboration consists of 
six imaging atmospheric Cherenkov telescopes (IACTs) located on the 
Roque de los Muchachos on the Canary island of La Palma, at 2200~m 
above sea level. 
A prototype telescope (CT1) started operation in 1992 and has undergone 
significant hardware upgrades since then. This telescope continues to operate as 
an independent instrument. The stereoscopic system of Cherenkov telescopes 
consists of five telescopes (CT2 - CT6), and has been taking data since 1996, 
initially with three and four telescopes, and since 1998 as a complete 
five-telescope system. Four of the telescopes (CT2, CT4, CT5, CT6) 
are arranged 
in the corners of a square with roughly 100 m side length, and one 
telescope (CT3)
is located in the center of the square. During 1997, when the data 
discussed in this paper were taken, CT2 was still used as stand alone detector.

The telescopes have an 8.5 m$^2$ tessellated reflector, focusing the 
Cherenkov light onto a camera with 271 photomultipliers (PMTs), covering 
a field of view of $4.3^\circ$ in diameter. A telescope is triggered when the 
signal in at least two adjacent PMTs exceeds an amplitude of 10 
(before June 1997)
or 8 (after  June 1997)
photoelectrons; in order to trigger the CT system and to initiate the readout 
of data, at least two telescopes have to trigger 
simultaneously. Typical trigger 
rates are in the 10-16 Hz range. The PMT signals are digitized and recorded 
by 120 MHz Flash-ADCs. The telescope hardware is
described by Hermann (\cite{her:95}); the trigger system and its performance are 
reviewed  by Bulian et al.\ (\cite{bul:98}).

\subsection{Reconstruction of air showers with the HEGRA IACT system}

The routine data analysis (see Paper 1 for details)
includes a screening of data to exclude data sets 
taken at poor weather conditions or with hardware problems. In particular, the 
mean 
system trigger rate proved to be a sensitive diagnostic tool. Reconstruction of 
data involves the deconvolution of Flash-ADC data 
(Hess et al.\ \cite{hess:98}), the 
calibration and flat-fielding of the cameras, the determination of Hillas 
image parameters, and the reconstruction of geometrical shower parameters 
based on the stereoscopic views of the air shower obtained with the different 
telescopes (Daum et al.\ \cite{daum:97}, Aharonian et al.\
\cite{ahar:97a}). 
The characteristic angular 
resolution for individual $\gamma$-rays is $0.1^\circ$; by sophisticated 
procedures $\gamma$-ray sources can be 
located with sub-arcminute precision (P\"uhlhofer et al.\ \cite{pul:97}). 

The separation of hadronic and electromagnetic showers   
is based on the shape of 
the Cherenkov images, in particular using the {\em width} parameter. The 
width of each image is normalized to the average {\em width} of a $\gamma$-ray 
image for a given impact parameter of the shower relative to the telescope, 
and a given image intensity. Here, impact distances are obtained from the 
stereoscopic reconstruction of the shower geometry. Cuts are then applied to 
the {\em mean scaled width} obtained by averaging the scaled width values 
over telescopes. For a point source both the {\em pointing} information and the 
image {\em shape} information are used. Each of them 
provides a cosmic-ray background rejection of up to 100.

The reconstruction of the energy of air showers is based on the 
relation between the shower energy and the image intensity ({\em size}) at a 
given distance from the shower axis (Aharonian et al.\ \cite{cell}).
This relation is tabulated based on 
Monte Carlo simulations, with the zenith angle of the shower as an 
additional parameter. The distance between a given telescope and the shower 
core is known from the stereoscopic reconstruction of the shower, with a 
typical precision of 10 m or less,
for not too distant showers. The energy estimates from the different 
telescopes are then averaged,
taking into account the slightly different sensitivities of
the telescopes. These sensitivities are calibrated to 1\%
by comparing the light yield in two telescopes for events with cores
halfway between the two telescopes (Hofmann \cite{hof:97}).
According to Monte Carlo simulations, this 
energy reconstruction provides an energy resolution of 15\% to 20\%, 
depending on the selection of the event sample. 
\subsection{Monte Carlo simulations of air showers and of the telescope 
response}

Any quantitative analysis of IACT data has to rely on detailed Monte Carlo 
simulations to evaluate the detection characteristics of the instrument.

The simulation of air showers and of Cherenkov light emission 
(Konopelko et al.\ \cite{kon:99}) includes all 
relevant elementary processes. On their trajectory to the detector, photons may be 
lost by ozone absorption, Mie scattering, and  Rayleigh scattering. Atmospheric
density profiles, ozone profiles and aerosol densities have been checked 
against local experimental data where available 
(e.g.\ Hemberger \cite{mark:98}).

On the detector side, the simulations include the wavelength dependence of 
the mirror reflectivity, of the light collection system, and of the PMT 
quantum efficiency. The point spread function of the mirror system is 
modeled after measurements of images of bright stars. The readout 
electronics is simulated in significant detail. PMT output waveforms are 
modeled by superimposing the response to single photoelectrons, with their 
relative timing and amplitude smearing. These signals are then
sampled, quantitized, and fed into the same analysis path as regular 
Flash-ADC data. The simulation includes the measured saturation effects both in 
the PMT/preamplifier and in the Flash-ADC.
Details concerning the Monte Carlo simulation  used here,
the performance of the system, and the comparison with experimental 
data are described by Konopelko et al.\ (\cite{kon:99}).

\section{The Mrk 501 data sample}

The analysis is based on the same data sample as used in Paper 1, 
corresponding to a total exposure time of 110~h, between March and 
October of 1997. 
Mrk 501 was observed in the so-called wobble mode, with the source 
positioned $\pm 0.5^\circ$ in declination
away from the optical axis of the telescopes, 
alternating every 20 min. For background subtraction, an equivalent region 
displaced by the same amount in the opposite direction is used. The 
separation of $1^\circ$ of these on-source and off-source regions is large 
compared to the angular resolution of the telescope system. In total, the 
sample comprises about 38,000 $\gamma$-ray events.

To select $\gamma$-ray candidates, the same loose cuts were applied as in 
Paper 1. In particular, the reconstructed shower direction had to be within 
$0.22^\circ$ from the source, and the {\em mean scaled width} parameter
had to be less than 
1.2. Events were accepted up to a maximum impact parameter of 200~m from 
the central telescope CT3. Events with larger impact parameters frequently 
suffer from truncated images in the cameras and, due to small angles 
between the stereo views, have larger uncertainties on the shower parameters. 
Since the analysis presented in the following emphasizes the control of 
systematic 
errors, it was felt that in this case the advantage of 
having clean events and a well-defined effective area at large energies --  
basically all events within the 200~m radius trigger above a few TeV -- 
outweighs the gain in statistics which could have been achieved by accepting all 
events \footnote{Replacing the 200~m by a 300~m restriction, 
the number of events reconstructed between 5 TeV and 10 TeV, and above 10 TeV 
increases by 20\% and 35\% after $\gamma$-ray selection cuts, respectively.}.
For the following analysis a software threshold of two or
more triggered telescopes, each with at least 40 recorded
photoelectrons was used.

Properties of this Mrk 501 data sample were examined in detail in Paper 1.
\section{Determination of energy spectra and sources of systematic errors}

Compared to single IACTs, stereoscopic IACT systems permit drastic reduction
of systematic errors, in particular for tasks like the precision
determination of energy spectra. Using the redundant information provided
by the multiple views, essentially all relevant characteristics, such as
the radial distribution of Cherenkov light or the trigger probabilities of
the telescopes can be verified experimentally 
(see also Hofmann \cite{hof:97}).
Simultaneous sampling of the intensity of the Cherenkov light front in
different locations emphasizes the calorimetric nature of the energy
determination, and reduces the effect of local fluctuations. Finally,
given the unambiguous reconstruction of the shower geometry and
the fact that at energies of one TeV virtually all events within 100~m
from the central telescope trigger the system, and that above a few TeV
almost all events within 200~m trigger, the effective detection area above 
1 TeV can be basically defined by pure geometry, 
without relying on simulations.
Only the threshold region requires a critical consideration. 

Under ideal conditions, the differential energy 
spectrum of the incident radiation is 
determined as 
\begin{equation}
\phi(E) = {r(E) \over \eta(E) A(E)} \, ,
\label{eqa}
\end{equation}
where $r(E)$ is the measured rate of $\gamma$-rays of energy $E$
after background subtraction, $A(E)$ 
is the effective detection area and $\eta(E)$ is the efficiency of the cuts 
applied to isolate the signal and to suppress the background.
The effective area and the cut efficiencies are usually derived  
from Monte Carlo simulations. 

A complication arises from the finite energy resolution, described by the 
response function ${\cal R}(E',E)$, the probability that a $\gamma$-ray
of energy $E$ is 
reconstructed at an energy $E'$. The measured rate is hence given by the 
convolution
\begin{equation}
r(E') = \int dE~{\cal R}(E',E)~\eta(E) A(E) \phi(E) \, .
\label{eqb}
\end{equation}
Eq.~\ref{eqb} can no longer be trivially inverted to yield $\phi(E)$. 
Options to 
find $\phi(E)$ include the explicit deconvolution using a suitable algorithm, 
which will usually make some assumption concerning the smoothness of 
the spectrum. Another approach is to assume a certain functional form for the
shape of the spectrum, and to determine free parameters such as the flux and 
the spectral index from a fit of Eq.~\ref{eqb} to the data. 
Finally, one can absorb 
the effect of the energy smearing into a modified effective area $A$, defined 
such that Eq.~\ref{eqa} holds. The latter approach is the simplest, but has the 
disadvantage that now $A$ depends on the assumed shape of the spectrum. 
However, with the $<20\%$ energy resolution provided by the HEGRA CT 
system, shape-dependent corrections are negligible for most practical 
purposes, and results are stable after
one iteration. Therefore, while both other techniques were pursued, the final 
results are based on this third method.

\begin{figure}
\resizebox{7.8cm}{!}{\includegraphics{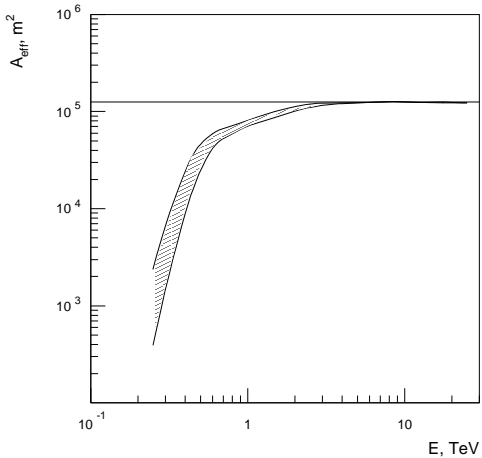}}
\caption
{Effective area of the HEGRA IACT system as a function of energy, 
for vertical $\gamma$-rays. The saturation at $10^5$ m$^2$ at high energies
reflects the cut in impact distance at 200~m relative to the central
telescope. The uncertainty of the effective area in the threshold region
caused by a 5\% variation of the detection threshold and by 
the interpolation in zenith angles is shown by the hatched area.
The corresponding systematic error on the effective area is 
$\sim 50$\% at 500 GeV and $\sim 10$\% at 1 TeV.} 
\label{fig_area}
\end{figure}

The typical energy dependence of the effective area is shown in 
Fig.~\ref{fig_area}. 
Below 1 TeV, the effective detection area rises steeply with 
energy, and then saturates at around $10^5$ m$^2$. The saturation reflects 
the cut on a maximum distance from the central telescope of 200~m.
The technical implementation of the energy reconstruction and 
flux determination
is described in detail in Paper 1. Compared to 
the brief discussion given above, a main complication for real data arises from
the dependence of $A$ on the zenith angle
$\theta$, which varies during runs and 
from run to run. In order to be able to interpolate between 
Monte Carlo-generated effective areas at certain discrete zenith angles, a semi-empirical 
scaling law is exploited, which relates the 
effective areas at different zenith 
angles. The variation of zenith angles with time is accounted for 
through replacing 
Eq.~\ref{eqa} by the sum over all events recorded within the observation
time $T$
\begin{equation}
\phi(E) = {1 \over T} \sum_{events} {1 \over \eta(E,\theta) A(E,\theta)}
\label{eqc}
\end{equation}
where each event is weighted with the appropriate effective area, given its 
energy and zenith angle. Note that for each period of a certain 
hardware configuration a set of effective areas is used which has been
determined from the Monte Carlo simulations which model in detail the
specific hardware performance, i.e.\ which take into account the 
trigger configuration and the mirror point spread function
(see Paper 1).

The key aspect in a precise and reliable determination of $\gamma$-ray 
spectra is the control of systematic errors. Sources of systematic errors 
include, e.g.,
\begin{itemize}
\item Systematic errors in the determination of the absolute energy scale.
\item Deviations from the linearity of the energy reconstruction, caused e.g. 
by threshold effects 
at very low energies, and possible saturation effects in the 
PMTs or the electronics at very high energies.
\item Systematic errors in the determination of the effective area $A$; 
particularly critical is the threshold region, where $A$ is a very steep 
function of $E$.
\item Systematic errors in the determination of the (energy-dependent) 
efficiency of the angular and image shape cuts.
\end{itemize}
As discussed in detail in Appendix A, a non-accurate modeling
of the detector response or the atmospheric transmission
possibly results in 1) a shift in the energy scale of the 
reconstructed $\gamma$-ray spectrum, and 
2) a distortion of the shape of the spectrum. 
Most systematic uncertainties, e.g.\ mirror reflectivities 
and PMT quantum efficiencies, 
exclusively contribute to an error of the energy scale.
The calculations in Appendix A show that the curvature of the spectrum
is reconstructed correctly provided that the Monte Carlo simulations 
accurately model the correlation between the detector threshold 
and the reconstructed energies including the fluctuations involved.
If the Monte Carlo description of this correlation is incorrect, the 
energies reconstructed in data and the effective areas computed from 
Monte Carlo simulations do not match. 
A shift of the reconstructed energies by a factor 
$(1+\epsilon)$ relative to the effective area $A$ used 
for the evaluation of the spectrum results in a 
flux error of
\begin{equation}
{\Delta \phi \over \phi} \approx {\Delta A(E) \over A(E)}
\approx \epsilon~{\mbox{d}\ln{A(E)} \over \mbox{d}\ln{E}}~~~.
\label{eqd}
\end{equation}
The flux error is potentially large in the threshold region, 
where $A$ varies quickly with $E$, $A \propto E^\beta$ 
with $\beta \simeq 6$ and $\Delta \phi / \phi
\approx 6 \epsilon$, but is negligible at high energies, where $A$ is
constant, and is simply governed by the geometrical cuts 
(see Fig.\ 1).

The remainder of this section is dedicated to a discussion of the
various sources of systematic errors.
\subsection{Reliability of the determination of the shower energy}

\begin{figure}
\resizebox{\hsize}{!}{\includegraphics{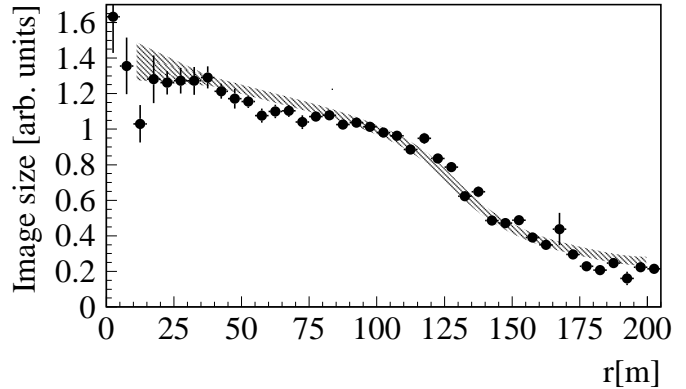}}
\caption
{Measured light yield as a function of distance to the shower core
for $\gamma$-ray showers in the energy range of about 0.9 to 1.8~TeV 
compared with Monte Carlo simulations (hatched band). 
Note that the light yield 
assigned to images depends on the field of view of the camera, and
on the definition of ``image'' pixels; the data should not be compared
to ``raw'' simulations not including such effects.}
\label{fig_yield}
\end{figure}

A crucial input for the determination of shower energies is of course the 
expected light yield as a function of core distance. Since the average core 
distance varies significantly with energy, inadequacies in the assumed 
relation will not only worsen the energy resolution, but will systematically 
distort the spectrum. With the redundant information
provided by a system of Cherenkov telescopes, it is possible to actually 
measure the light yield as a function of the distance from the shower core 
and to verify the simulations (Aharonian et al. \cite{ahar:98b}). 
Briefly, the idea is to 
select showers with a fixed impact parameter relative to a telescope A, 
and with a fixed light yield in this telescope. This provides a sample of 
showers of constant energy, and now the light yield in other telescopes can 
be measured as a function of core distance. Fig.~\ref{fig_yield} shows the 
characteristic shape of the light pool for $\gamma$-ray showers, which is 
well reproduced by the simulation; this holds also for the variation of the 
shape with shower energy and zenith angle 
(Aharonian et al. \cite{ahar:98b}).

\begin{figure}
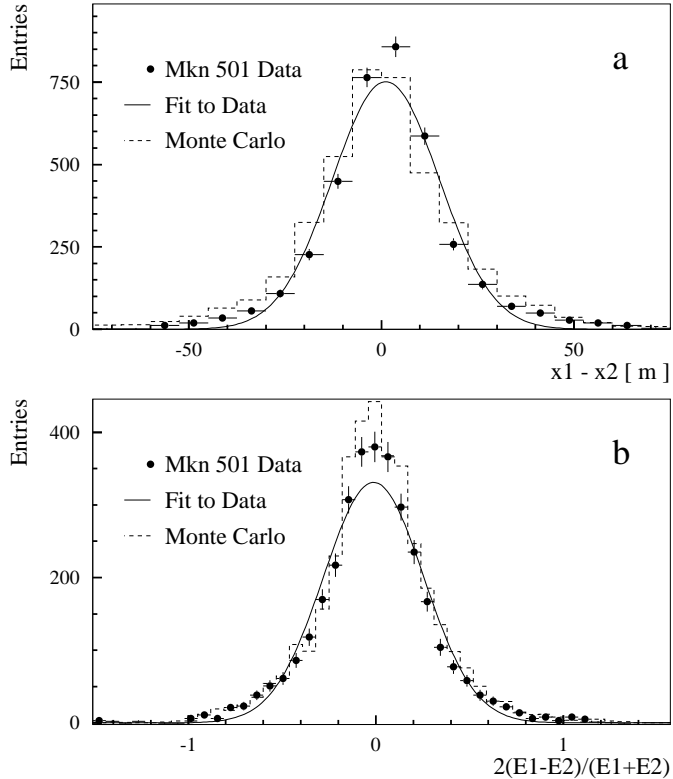

\resizebox{\hsize}{!}{\includegraphics{501.core_a.epsi}}\\[1ex]
\resizebox{\hsize}{!}{\includegraphics{501.core_b.epsi}}
\caption
{(a) Difference in the $x$ coordinate of the shower core as measured
by independent subsystems of two telescopes for events where all four
telescope triggered. Showers are selected to provide a minimum stereo
angle of $20^\circ$ in each subsystem, and a maximum core distance of
200~m from the central telescope. Points show the data, the dashed line
the simulation, and the full line a Gaussian fit to the data with a 
width of 14~m. (b) Comparison of the energies measured by the two
subsystems. The width of the Gaussian fit is 25\%. Same cuts as in (a).}
\label{fig_res}
\end{figure}

A second key ingredient in the energy determination is the reconstruction of 
the core location. Unlike in the case of the angular resolution, where it is 
easy to show that the simulation reproduces the data by comparing the 
distribution of shower directions relative to the source with the simulations, 
a direct check of the precision of the core reconstruction is not possible. 
However, noting that two telescopes suffice for a stereoscopic 
reconstruction, one can split up the four-telescope system into two systems 
of two telescopes and compare the results (Hofmann \cite{hof:97}). 
Figure ~\ref{fig_res}a illustrates 
the difference in core coordinates bet\-ween the two subsystems for data
and for the Monte Carlo simulations. 
As can be recognized 
the Monte Carlo simulations accurately
predict the distribution of the distances between the two cores.
Under the assumption that the two measurements are uncorrelated and that 
the reconstruction accuracy achieved with two telescopes is the same for
two telescope and four telescope events
the width of the difference distribution should be $\sqrt{2}$ 
times the resolution of a two-telescope system.
By this means, the 2 telescope resolution is determined to be 
$14~{\rm m}~/~\sqrt{2}~\approx~10$~m.
The Monte Carlo simulations show, that this reconstruction accuracy  
does indeed agree nicely with the true reconstruction 
accuracy for 2 telescope events.

The same technique can be used to study the energy resolution
(Fig.~\ref{fig_res}b; see also Hofmann \cite{hof:97}). 
Also here the data is in excellent agreement with the simulations.
In this case, the Monte Carlo studies predict, that the energies measured 
with the two subsystems are considerably correlated:
assuming uncorrelated estimates, an energy resolution of 
$25\%~/~\sqrt{2}~=~ 18\%$ is inferred;
the Monte Carlo simulations predict a true energy resolution for two
telescope events of 23\%.
On the basis of simulations, fluctuations in the shower height can be 
identified as the origin of this correlation. Work is ongoing to use the 
stereoscopic determination of the shower height to improve the energy resolution. 
The excellent agreement between data and Monte Carlo simulations confirms 
that the experimental effects entering the energy determination are well under control, 
and that consequently the estimate of the energy resolution based on the simulations 
is reliable.
\subsection{The absolute energy scale}
The absolute energy calibration of IACTs is a significant challenge, lacking a 
suitable monoenergetic test beam. Factors entering the absolute energy 
calibration are the production of Cherenkov light in the shower, the 
properties and the transparency of the atmosphere, and the response of the 
detection system involving mirror reflectivities, PMT quantum efficiencies, 
electronics calibration factors etc. So far, mainly three techniques were used 
to calibrate the HEGRA telescopes:
\begin{itemize}
\item The comparison between predicted and measured cosmic-ray detection 
rates. Given the integral spectral index of 1.7 for cosmic rays, an error 
$\epsilon$ in the energy scale results in an error of $1.7 \epsilon$ 
in the rate above a given threshold. 
Apart  from the slightly different longitudinal 
evolution of hadronic and of $\gamma$-ray induced showers, this test 
checks all factors entering the calibration. The Monte Carlo simulations
reproduce the measured cosmic-ray trigger rates within 10\% (Konopelko et 
al. \cite{kon:99}). The IACT  system was furthermore used to determine 
the flux of 
cosmic-ray protons in the 1.3 to 10~TeV energy range
(Aharonian et al.\ \cite{ahar:98c}). 
The measured flux of protons
\begin{equation}
{\rm d} N_{\rm p}/{\rm d} E=(0.11 \pm 0.02_{\rm stat} \pm 0.05_{\rm sys}) 
\times
\end{equation}
\[
\hspace*{0.5cm}
E^{-2.72 \pm 0.02_{\rm stat} \pm 0.15_{\rm sys}}~~~
\rm s^{-1}sr^{-1} m^{-2} TeV^{-1}
\]
excellently agrees with a fit to the combined data of all
other experiments, indicating that systematics are well
under control and that the assigned systematic 
errors -- partially related to the
energy scale -- are rather conservative.
\item Given the measured characteristics of the telescope components such 
as mirrors or PMTs, the sensitivity of the telescopes can be determined with 
an overall error of 22\%.
\item Using a distant, calibrated, pulsed light source, an overall calibration 
of the response of the telescope and its readout electronics could be 
achieved, with a precision of 10\% (Fra{\ss} et al. \cite{frass:97}).
\end{itemize}
The last three techniques have to rely on the Monte Carlo simulations of the 
shower and of the atmospheric transparency. While Monte Carlo simulations 
have converged and different codes produce consistent results, details of the 
atmospheric model and assumptions concerning aerosol densities can 
change the Cherenkov light yield on the ground by about 8\% 
(Hemberger 1998). 

Within their errors, all calibration techniques are consistent. 
Overall, we believe that a 15\% systematic error on the energy scale is 
conservative, given the state of simulations and understanding of the instrument. 
Based on the comparison with cosmic-ray rates, 
one would conclude that the actual calibration
uncertainty is below 10\%.
\subsection{The threshold region and associated uncertainties}

\begin{figure}
\hspace*{0.5cm}
\resizebox{6.4cm}{!}{\includegraphics{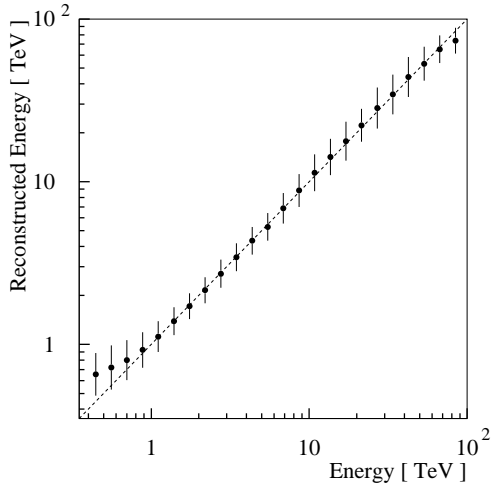}}
\caption
{Mean reconstructed shower energy as a function of the true energy for
showers incident under 20$^\circ$ zenith angle,
also showing the rms errors of the energy reconstruction.}
\label{fig_ereco}
\end{figure}

As discussed earlier in detail, the threshold region is very susceptible
to systematic errors. In the sub-threshold regime events trigger only
because of upward fluctuations in the light yield, and energy estimates
tend to be biased towards larger values. Fig.~\ref{fig_ereco} shows the
mean reconstructed energy as a function of the true energy for a sample
of simulated events at typical zenith angles.
Note that for the 1997 data set we have a noticeable  
number of detected $\gamma$-rays events below 500 GeV. 
However in this energy region 
the bias is so strong that a reliable correction is no 
longer possible. Therefore spectra will only be quoted above this energy.
\begin{figure}
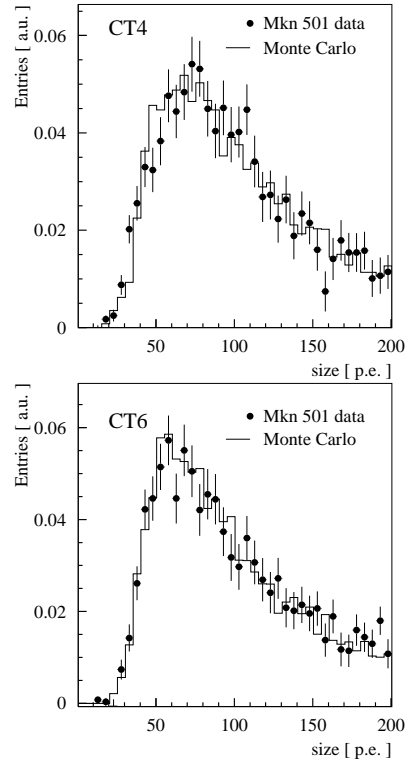

\hspace*{1.5cm}
\resizebox{5.2cm}{!}{\includegraphics{501.size_a.epsi}}\\
\hspace*{1.5cm}
\resizebox{5.2cm}{!}{\includegraphics{501.size_b.epsi}}
\caption
{{\em Size}-distributions of $\gamma$-ray induced air showers 
for data and simulations for two of the four telescopes.
The data is from a period of a certain trigger configuration,
namely ``Data-period I'' of Paper 1, and the Monte Carlo simulations used
the telescope specific single pixel trigger probability as function
of signal amplitude derived from this data 
(Mkn 501 data, background subtracted, Monte Carlo weighted 
according to the results in Sect.\ 5).}
\label{fig_size}
\end{figure} 

The first major source of systematic errors at low energies is the 
detailed modeling of the triggering process. In the simulation great
emphasis was placed on the correct description of the pixel trigger
probabilities as a function of signal amplitude.
For each trigger configuration and for each telescope this dependence 
has been derived from air shower data using the recorded information 
about which pixels of a triggered telescope surpassed the 
discriminator threshold and which pixels did not.
As discussed above and in Appendix A, 
the onset of the {\em size}-distribution provides
a sensitive test of the quality of the simulation. Fig.~\ref{fig_size}
shows the measured and simulated {\em size}-distribution for two telescopes
for a certain trigger configuration (``Data-period I'' of Paper 1).
For our current best simulations, fits to the rising edges of the 
{\em size}-distributions indicate for the individual telescopes and the 
different trigger configurations deviations of the {\em size} scale between
data and Monte Carlo on the 5\% level.
For conservatively estimating the systematic error on the shape of the 
Mkn 501 spectrum, we allow for a $\pm$5\% correlated shift of the 
thresholds of all four telescopes.
The resulting uncertainties in the effective detection area $A$ are indicated 
in Fig.~\ref{fig_area}; as expected, $A$ significantly changes 
in the threshold regime, but remains constant well above threshold.

Random threshold variations (Gaussian-distributed with a width of 15\%)
between individual pixels were found to be of
relatively small influence compared to the systematic threshold shifts.

As another source of systematic errors of special importance in the
threshold region, the accuracy of the 
energy reconstruction for zenith angles between the discrete 
simulated zenith angles ($\theta=0^\circ$, $20^\circ$, $30^\circ$, $45^\circ$) 
has been considered.
Imperfections in the scaling law used to relate the Cherenkov light yield
at different zenith angles $\theta$ could result in a systematic 
shift of the reconstructed energy at intermediate values of $\theta$. 
To test the description, the energy of the Monte Carlo showers with $\theta = 30^\circ$
was determined with the light yield tables of the $0^\circ$- and $45^\circ$-showers
and the result was compared to the result based on using the $30^\circ$ light yield table.
The systematic shift of the reconstructed energies was about 5\% below 1 TeV and
2\% above 1 TeV; based on this and other studies we 
believe that using the full set of simulations,
systematic shifts in the reconstructed energy are below
5\% and 2\%, below and above 1 TeV respectively.

The modified effective area used for the determination of the spectrum 
slightly depends on the assumed source spectrum. 
At the lowest and at the highest energies the source spectrum can not be
determined with high statistical accuracy. For energies below 1 TeV we conservatively
estimate the corresponding uncertainty in the modified effective area, by varying
the spectral index of an assumed source spectrum dN/dE$\propto$E$^{-\alpha}$ 
from $\alpha=1.5$ to $\alpha=3$. At the highest energies above 15 TeV we vary the
assumed source spectrum from a broken power law to a power law with an exponential cutoff, 
both specified by fits to the data. In addition we explore the statistical
significance of the $\gamma$-ray excess at the highest energies by a dedicated 
$\chi^2$-analysis (see Sect.\ \ref{RES}).

In the threshold region of the detector
the total systematic error is dominated
by the systematic shift of the telescope thresholds 
and by the possible systematic shift in the reconstructed 
energy due to the zenith angle interpolation.
The total systematic error is computed by summing up the 
individual relative contributions in quadrature.
\subsection{Saturation effects at high energies}
\begin{figure}
\hspace*{0.5cm}
\resizebox{6.4cm}{!}{\includegraphics{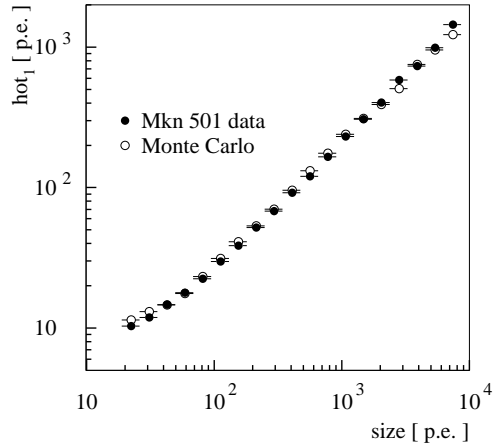}}
\caption
{The mean signal in the peak pixel, hot$_1$, as a function of the image 
{\em size}, for data (full points) and simulations (open points).}
\label{fig_ratio}
\end{figure} 
The PMTs and readout chains of the HEGRA system telescopes provide a linear 
response up to amplitudes of about 200 photoelectrons. At higher intensities, 
nonlinearities of the PMT become noticeable, and also the 
8-bit Flash-ADC saturates. With typical signals in the highest pixels of 25
photoelectrons per TeV $\gamma$-ray energy,
the effects become important for energies around 10 TeV and above.

Saturation of the Flash-ADC can be partly recovered by using the recorded 
length of the pulse to estimate its amplitude, effectively providing a 
logarithmic characteristic (Hess et al.\ \cite{hess:98}). 
The saturation characteristics 
of the PMT/preamplifier assembly have been measured, and a correction is 
applied. These nonlinearities and the Flash-ADC saturation are included in 
the simulations. 
As a very sensitive quantity to test the handling of 
saturation characteristics, the dependence of the pulse height 
in the peak pixel on the 
image {\em size} emerged (Fig.~\ref{fig_ratio}).
Data and simulations are in very good agreement up to pixel amplitudes
exceeding 10$^3$ photoelectrons, equivalent to $>$50 TeV $\gamma$-ray showers.
Older versions of the simulations, 
which did not properly account for PMT/preamp nonlinearities, 
showed marked deviations for {\em size}-values above 10$^3$. 

Independent tests of saturation and saturation corrections were provided
by omitting, both in the data and in the simulation, the highest pixels
in the image, and by comparing event samples in different ranges of 
core distance and zenith angle. The saturation effect is non-negligible
only at very high energies, namely $E \geq 15 \, \rm TeV$.  
However, all tests show that given the quality of our current simulations, 
systematic errors induced by saturation effects even in this energy region 
are yet small compared to the statistical errors.
\subsection{Efficiency of cuts}

Among the sources of systematic errors, the influence of cuts
\begin{figure}
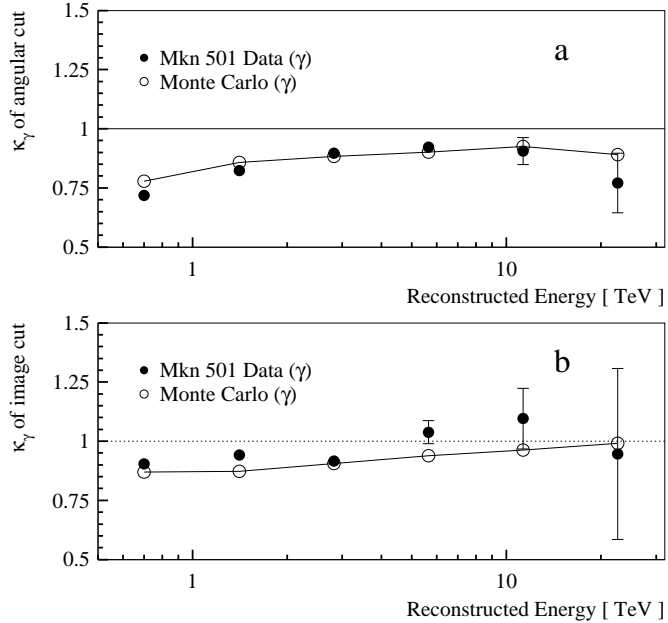

\resizebox{\hsize}{!}{\includegraphics{cut_eff.epsi}}\\
\resizebox{\hsize}{!}{\includegraphics{cut_effb.epsi}}
\caption
{Efficiency of the cut on the shower direction relative to 
the Mrk~501 location 
after the software threshold of at least 40 photoelectrons 
in two or more telescopes (a),
and of the shape cuts used to enhance $\gamma$-rays (b), as a function of
energy. Full points show the measured efficiencies, open points  
the results of the Monte Carlo simulations. 
Efficiencies determined from data can be 
larger than one due to background fluctuations.}
\label{fig_cuteff}
\end{figure} 
 is less critical. The large flux of 
$\gamma$-rays from Mrk 501 combined with the excellent background 
rejection of the IACT system allows to detect the signal essentially without 
cuts. In the analysis, one can afford to apply only rather loose cuts, which 
keep over $80\%$ of all $\gamma$-rays; only at the lowest energies,
a slightly larger fraction of events is rejected. 
Since only a small fraction of the 
signal is cut, the uncertainty in the cut efficiency is a priori small; in 
addition, the efficiency can be verified experimentally by comparing with the 
signal before cuts, see, e.g., Fig.~\ref{fig_cuteff}.
The cut efficiencies in Fig.~\ref{fig_cuteff} 
deviate from the results shown in Paper 1 on the
5\%-level due to slightly improved Monte Carlo simulations.
We conservatively estimate the systematic error the cut efficiencies, 
to be 10\% at 500 GeV decreasing to a constant value of 5\% at 2 TeV 
and rising above 10 TeV to 15\% at 30 TeV. 
At low energies the acceptances are corrected according to the 
measured efficiencies.
\subsection{Other systematic errors and tests}

The precision in the determination of the effective areas is partly limited by 
Monte Carlo statistics. The universal scaling law used to relate  Monte Carlo 
generated effective areas at different zenith angles involves a rescaling of 
shower energies. Statistical fluctuations in a Monte Carlo sample at a given 
energy and angle will hence influence the area  over a range of 
energies. Because of the resulting slight correlation, Monte Carlo statistics is 
in the following included in the systematic errors.

To test for systematic errors, the data sample was split up into subsamples 
with complementary systematic effects, and spectra obtained for these 
subsamples were compared. Typical subsamples include
\begin{figure}
\resizebox{\hsize}{!}{\includegraphics{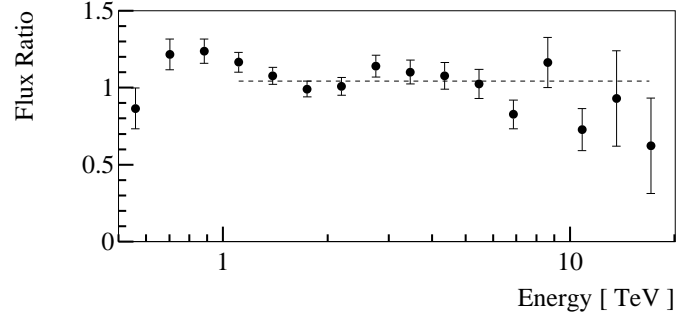}}
\caption
{Ratio of the spectra computed with the events with small (0 to 120~m) and with 
large (120 m to 200 m) impact distances relative to the central telescope. A fit to a
constant gives a ratio of $1.04 \pm 0.02$ with a 
$\chi^2$ of 23.1 for 12 degrees of freedom.
The fit is restricted to the energy region of reasonable systematic errors, 
i.e. with an effective detection area for both event samples larger than 
$\sim\,10^4$ m$^2$.
}
\label{fig_subsamplesb}
\end{figure} 
\begin{itemize}
\item Events where 2, 3 or 4 telescopes are used in the reconstruction.
\item Events passing a higher `software trigger threshold', requiring
e.g. a minimum {\em size} of 100 photoelectrons, or a signal in the
two peak pixels of at least 30 photoelectrons.
\item Events with showers in a certain distance range from the center
of the system (CT3), e.g. 0-120~m compared to 120-200~m. This comparison tests 
systematics in the light-distance relation as well as the correction of 
nonlinearities in the telescope response.
\item Events where all pixels are below the threshold for nonlinearities.
\item Events in different zenith angle ranges. 
The comparison of these spectra provides a sensitive test of the entire 
machinery, and also of nonlinearities, where the data at larger angles
should be less susceptible because of the smaller {\em size} at a given 
energy.
\end{itemize}
For each of the subsamples, the effective area and cut efficiencies 
were determined, and a flux was calculated. In all cases, deviations
between subsample spectra were insignificant, or 
well within the range of systematic errors.
Among the variables studied, the most significant indication of
remaining systematic effects is seen in the comparison 
of different ranges in shower impact parameter relative to 
the central telescope.
The ratio of the spectra determined with the data of 
small ($<$120 m) and large (between 120 m and 200 m) impact distances
is shown in Fig.\ \ref{fig_subsamplesb}. A fit to a constant 
gives a mean ratio of $1.04 \pm 0.02$ with a $\chi^2$-value of 23.1 for 12 
degrees of freedom, corresponding to a chance probability 
for larger deviations of 5\%.
\section{Experimental results}
\label{RES}
\begin{figure}
\resizebox{\hsize}{!}{\includegraphics{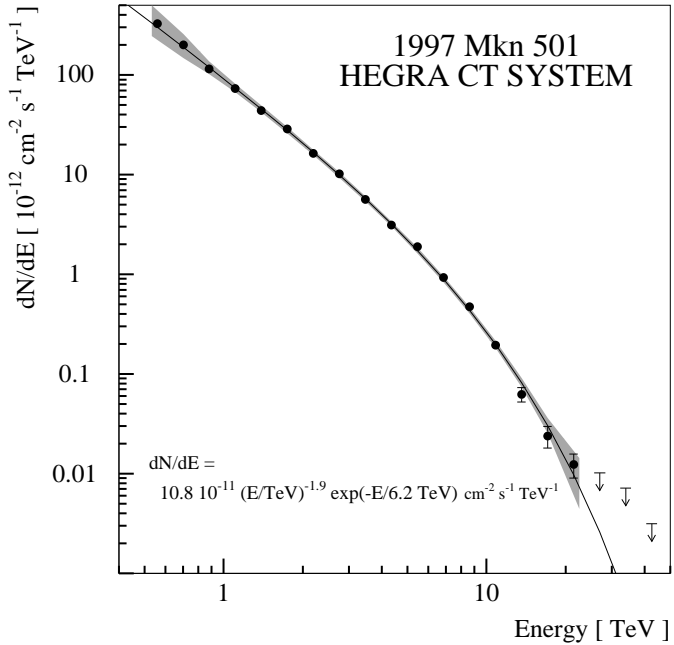}}
\caption
{Time-averaged energy spectrum of Mrk 501 for the 1997 observation
period. Vertical errors bars indicate statistical errors.
The hatched area gives the estimated
systematic errors, except the 15\% uncertainty on the absolute energy scale. 
The lines shows the fit discussed in the text.}
\label{fig_spectrum} 
\end{figure}

In this section we first present the 1997 Mkn 501 time averaged energy 
spectrum.
As discussed already in the introduction the derivation of a time 
averaged spectrum is meaningful since the changes in the spectral shape
during the HEGRA observations were rather small, i.e.\ they were too small
to be assessed with an accuracy of typically between 0.1 and 0.3 
in the diurnal spectral indices.
Moreover, as described in Paper 1, dividing the data into groups 
according to the absolute flux level or according to the rising or 
falling behavior of the source activity yielded mean spectra which did 
not differ significantly from each other in the one to ten TeV energy range.
The weakness of the correlation between the absolute flux and the spectral 
shape will further be substantiated below over the energy region from 
500~GeV to 15~TeV.
Nevertheless, the importance of the spectral constancy should not be 
overestimated. If the spectral variability is not tightly correlated 
with the absolute flux, diurnal spectral variability characterized 
by a change of the spectral index at several TeV
by approximately $\pm$0.1 is surely consistent with the HEGRA data.
The time-averaged energy spectrum is shown in Fig.~\ref{fig_spectrum}. 
For the determination of the spectrum also at energies below 800 GeV, 
only the data from zenith angles smaller 30$^\circ$ have been used 
(80 h observation time).
The measurements extend from 500 GeV to 24 TeV. 
The hatched region in Fig.\ \ref{fig_spectrum} ff.\
gives our estimate of the systematic errors on the shape of the spectrum,
except the 15\% uncertainty on the absolute energy scale.
The spectrum shows a gradual steepening over the entire energy range. 
A fit of the data from 500 GeV to 24 TeV 
with a power law model with an exponential cut off gives:
\begin{equation}
\label{pexp}
dN/dE \, = \, N_0 \,(E/1\, {\rm TeV})^{-\alpha}\,\exp{(-E/E_0)}~~,
\end{equation}

\noindent
$N_0=(10.8 ~\pm0.2_{\rm stat}~\pm2.1_{\rm sys}) 
\cdot 10^{-11} \, \rm cm^{-2} s^{-1} TeV^{-1}$,   
$\alpha=1.92 ~\pm0.03_{\rm stat} ~\pm0.20_{\rm sys}$, and 
$E_0=(6.2 ~\pm0.4_{\rm stat} ~(-1.5 ~+2.9)_{\rm sys})$ TeV.
The systematic errors on the fit parameters result from 
worst case assumptions concerning the 
systematic errors of the data points, 
and their correlations and include the error 
caused by the 15\% uncertainty in the energy scale. 
The errors on the fit parameters, especially on 
$\alpha$ and $E_0$, are strongly correlated.
The variation of only one of the parameters within the quoted error range 
yields spectra which are inconsistent with the measured spectrum.
The data points and their errors are summarized in Table~1.
\begin{table}
\caption
{The time-averaged differential spectrum of Mkn 501}
{\footnotesize
\begin{center}
\begin{tabular}{rrrr}
\hline
\\
{\small E$^a$}  & {\small ${\rm d}N/{\rm d}E$~~$^b$} &
   {\small $\sigma_{\rm stat}~~^c$} & {\small $\sigma_{\rm sys}~~~^d$} \\
\\
\hline 
\\
           0.56  &    3.29 $10^{-10}$ &     1.68 $10^{-11}$ &  (+1.82  -1.07) $10^{-10}$\\
           0.70  &    2.01 $10^{-10}$ &     7.45 $10^{-12}$ &  (+6.33  -4.75) $10^{-11}$\\
           0.88  &    1.15 $10^{-10}$ &     3.45 $10^{-12}$ &  (+1.88  -1.68) $10^{-11}$\\
           1.11  &    7.33 $10^{-11}$ &     1.88 $10^{-12}$ &  (+7.42  -7.08) $10^{-12}$\\
           1.39  &    4.40 $10^{-11}$ &     1.10 $10^{-12}$ &  (+3.75  -3.45) $10^{-12}$\\
           1.75  &    2.87 $10^{-11}$ &     7.25 $10^{-13}$ &  (+2.12  -1.97) $10^{-12}$\\
           2.20  &    1.64 $10^{-11}$ &     4.64 $10^{-13}$ &  (+1.03  -0.97) $10^{-12}$\\
           2.76  &    1.02 $10^{-11}$ &     3.14 $10^{-13}$ &  (+6.71  -6.30) $10^{-13}$\\
           3.46  &    5.64 $10^{-12}$ &     1.98 $10^{-13}$ &  (+3.50  -3.30) $10^{-13}$\\
           4.35  &    3.12 $10^{-12}$ &     1.27 $10^{-13}$ &  (+1.94  -1.83) $10^{-13}$\\
           5.46  &    1.90 $10^{-12}$ &     8.73 $10^{-14}$ &  (+1.24  -1.16) $10^{-13}$\\
           6.86  &    9.29 $10^{-13}$ &     5.24 $10^{-14}$ &  (+6.40  -5.99) $10^{-14}$\\
           8.62  &    4.71 $10^{-13}$ &     3.28 $10^{-14}$ &  (+3.42  -3.19) $10^{-14}$\\
          10.83  &    1.95 $10^{-13}$ &     1.82 $10^{-14}$ &  (+1.69  -1.56) $10^{-14}$\\
          13.60  &    6.24 $10^{-14}$ &     1.02 $10^{-14}$ &  (+9.32  -8.11) $10^{-15}$\\
          17.08  &    2.39 $10^{-14}$ &     5.85 $10^{-15}$ &  (+5.87  -4.71) $10^{-15}$\\
          21.45  &    1.24 $10^{-14}$ &     3.35 $10^{-15}$ &  (+5.93  -4.01) $10^{-15}$\\
          26.95  &  $<$ 1.0 $10^{-14}$ $\,^e$&&\\
          33.85  &  $<$ 7.2 $10^{-15}$ $\,^e$&&\\
          42.51  &  $<$ 3.1 $10^{-15}$ $\,^e$&&\\
\\
\hline
\\
\multicolumn{4}{l}{\footnotesize  
$^a$ energy in TeV}\\
\multicolumn{4}{l}{\footnotesize  
$^b$ in (cm$^{-2}$ s$^{-1}$ TeV$^{-1}$)}\\
\multicolumn{4}{l}{\footnotesize  
$^c$ statistical error in (cm$^{-2}$ s$^{-1}$ TeV$^{-1}$)}\\
\multicolumn{4}{l}{\footnotesize  
$^d$ systematic error on the shape of the spectrum in}\\
\multicolumn{4}{l}{\footnotesize  
~~~(cm$^{-2}$ s$^{-1}$ TeV$^{-1}$)}\\
\multicolumn{4}{l}{\footnotesize  
$^e$ upper limits in (cm$^{-2}$ s$^{-1}$ TeV$^{-1}$) at 2 $\sigma$ confidence level}\\
\end{tabular}
\label{tab_results}
\end{center}
}
\end{table}
\begin{figure}
\resizebox{\hsize}{!}{\includegraphics{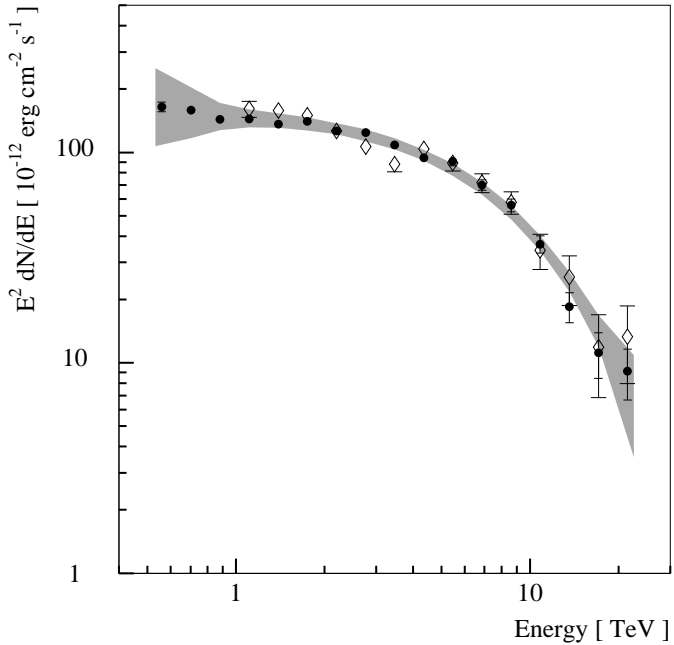}}
\caption
{The spectral energy distribution $E^2 \mbox{d}N/\mbox{d}E$, for
the data set of low zenith angles ($\theta<30^\circ$, full circles) 
and for the data set of large zenith angles 
($\theta$ between $30^\circ$ and 45$^\circ$, 32 h observation time, 
open symbols).
Since the observation periods do not overlap for the
variable source, the spectra are normalized at the energy 2 TeV.
The hatched band indicates the systematic error on the shape of the spectrum
for the low zenith angle data. The systematic error on the high zenith angle spectrum 
at energy E approximately equals the systematic error 
on the low zenith angle spectrum at energy E/2.
}
\label{fig_seng2}
\end{figure} 

\begin{figure}
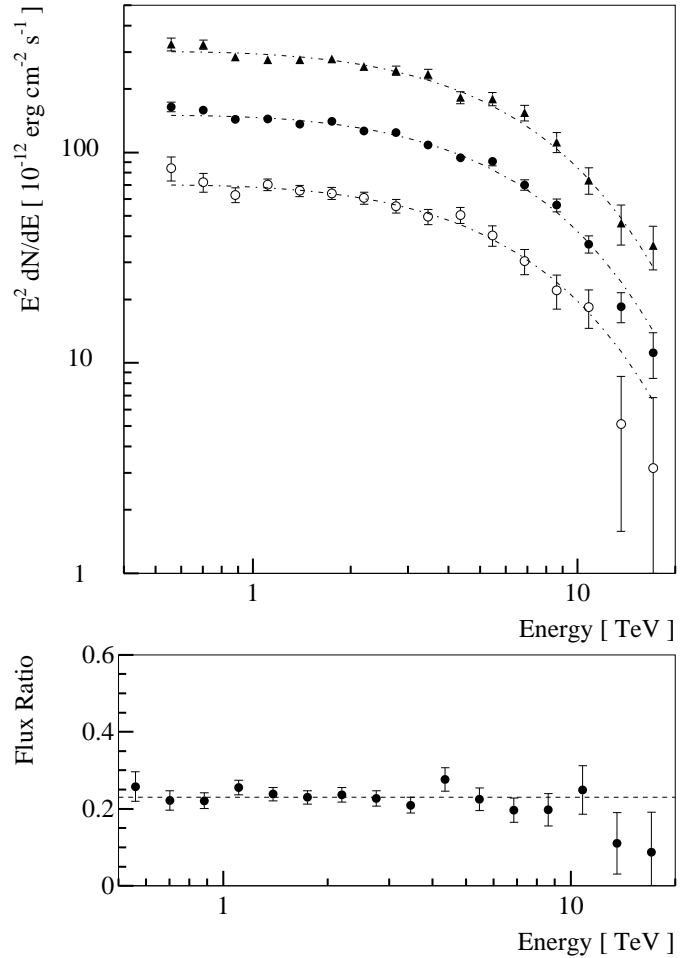

\resizebox{\hsize}{!}{\includegraphics{501.spec03b.epsi}}
\vspace*{2ex}
\resizebox{\hsize}{!}{\includegraphics{501.spec03a.epsi}}
\caption
{
The upper panel illustrates the spectral energy 
distribution $E^2 \mbox{d}N/\mbox{d}E$, for
the full data set (full circles), for periods of low flux
(open circles), and for periods of high flux (triangles) 
( dN/dE(2 TeV) above 30 and below 16 times $10^{-12} {\rm cm^{-1} s^{-1} TeV^{-1}}$). 
Only the statistical errors are given here;
the systematic errors enter the three spectra in the same way 
and can be neglected comparing the three spectra.
The dashed lines indicate the shape of the mean spectrum (fit from Eq.
\ref{pexp}) overlaid over all three spectra to simplify the comparison of the 
shape of the three spectra.
In the lower panel the ratio of the low flux spectrum divided 
by the high flux spectrum is shown. The dashed line gives the
fit to a constant. The $\chi^2$-value is 12.3 for 15 degrees of freedom.
}
\label{fig_seng}
\end{figure} 
\begin{figure}
\resizebox{\hsize}{!}{\includegraphics{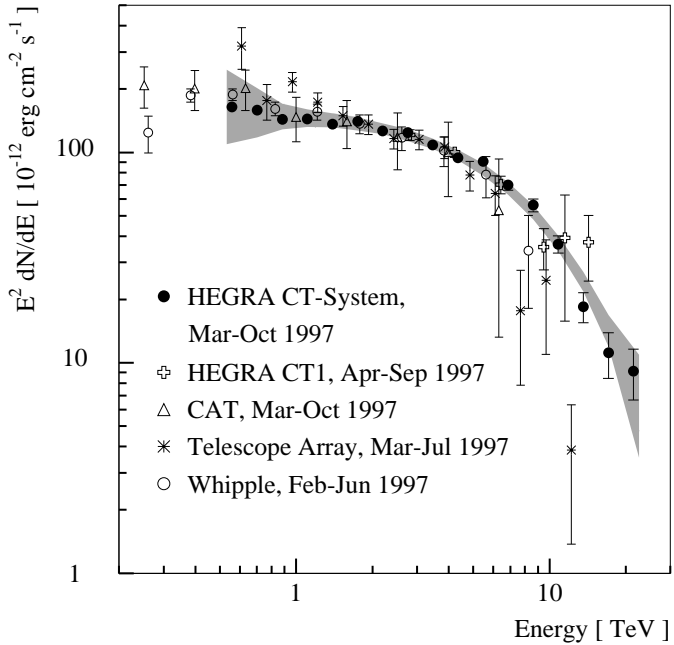}}
\caption
{The Time-averaged spectrum of Mrk 501 during 1997, compared with published
results from other experiments 
(Aharonian et al.\ \cite{Mkn501PartII},
Hayashida et al.\ \cite{hay:98}, Samuelson et al.\ \cite{samuel:98}, 
Barrau et al.\ \cite{barr:97}). 
Since the observation periods do not completely overlap for the
variable source, the spectra are normalized at the energy 2 TeV.
For the HEGRA system the hatched area shows the systematic errors on the
shape of the spectrum as described in the text.
For the other experiments only 
statistical errors are shown. 
Only data points with a signal to 
noise ratio larger than one have been used.
}
\label{fig_comp}
\end{figure} 
In the highest energy bin (19 TeV to 24 TeV) 40 excess events are
found above a background of 13 events, corresponding to a nominal
significance of 
S = (N$_{\rm on}$ - N$_{\rm off}$) / $\sqrt{\rm N_{on} + N_{off}}$  of
3.7 $\sigma$. 
However, due to the steep 
spectrum in this energy range, a part of these events may represent
a spill-over from lower energies. To provide an absolutely reliable
lower limit on the highest energies in the sample,
the spectrum was fit to the form of Eq.~\ref{pexp}, but 
with a sharp cutoff at $E = E_{cut}$:
$dN/dE = N_0$ $(E/1 ~\rm TeV)^{-\alpha}$ $\exp{(-E/E_0)}$ $\Theta(E_{cut}-E)$.
The best fit is achieved with
$E_{\rm cut} = 28$~TeV; the $2 \sigma$ lower limit is $E_{cut} = 16$~TeV.

Fig.\ ~\ref{fig_seng2} illustrates the spectral energy distribution, 
$E^2 dN/dE(E)$ as determined from the small zenith angle data
($<$30$^\circ$, energy threshold 500 GeV) 
and the large zenith angle data 
(30$^\circ$ to 45$^\circ$, 32 h observation time, energy threshold 1 TeV).
Note that the large zenith angle data has mainly been acquired during the 
second half of the 1997 data taking period. 
Nevertheless the shape of both spectra 
agrees within the statistical and systematic errors.
The combined small and large zenith angle data set yields the same
lower limit on $E_{cut}$ of 16~TeV as derived from the small zenith 
angle data alone. It can be recognized that the spectral 
energy distribution is essentially flat 
from 500 GeV up to $\simeq$ 2 TeV. 

Figure ~\ref{fig_seng} (upper panel) shows
the spectral energy distribution for the overall data sample
and for periods of high 
and low flux separately: dN/dE(2 TeV) determined on diurnal basis 
above 3 and below 1.6 $10^{-11} {\rm cm^{-1} s^{-1} TeV^{-1}}$,
with a ratio of the mean fluxes close to 5.
The high and low flux spectra agree within statistical errors, as 
shown by the ratio of both spectra, presented in Fig.~\ref{fig_seng} (lower panel).
The systematic error is to good approximation the same
for both data samples and cancels out in the ratio.
The result thus confirms our previous conclusion about the flux-independence
of the spectrum of Mkn 501 in 1997 between 1 and 10 TeV (Paper 1).
Now the statement is  extended to the broader energy region,
from 500 GeV to 15 TeV. 
From 1~TeV to several TeV the slope of the spectrum is determined 
with high statistical accuracy, e.g.\ a power law fit in the energy region 
from 1~TeV to 5~TeV gives a differential index of 
-2.23~$\pm0.04_{\rm stat}$ and -2.26 $\pm$0.06$_{\rm stat}$ 
for the high and the low flux spectrum respectively. 
In the narrow energy range from 500 GeV to 1 TeV the statistical 
uncertainty on the spectral index is considerably larger, we compute 
0.2 for the high flux sample and 0.4 for the low flux sample.
Therefore, our 1997 Mkn 501 data would not contradict a correlation
of emission strength and spectral shape below 1~TeV as tentatively 
reported by the CAT-group (Djannati-Atai et al.\ \cite{CATcorrelation}).

For completeness, the HEGRA IACT system data are plotted in Fig.~\ref{fig_comp} jointly 
with the HEGRA CT1 (Aharonian et al.\ \cite{Mkn501PartII}), 
the CAT (Barrau et al.\ \cite{barr:97}),
the Telescope Array (Hayashida et al. \cite{hay:98}), and
the Whipple (Samuelson et al.\ \cite{samuel:98}) results
concerning the Mrk 501 energy spectrum during the 1997 outburst.
Generally a good agreement can be recognized in the overlapping energy 
regions, except for a steeper Telescope Array spectrum.
\section{Discussion}

The observations of Mkn~501 by the HEGRA IACT system during the 
long outburst in 1997 convincingly demonstrate for the first time that 
the energy spectrum of the source extends well beyond 10 TeV. 
We believe that this very fact,
together with the discovery of a time- and flux- 
independent stable spectral
shape of the TeV radiation will have considerable impact on our
understanding of the nonthermal processes in AGN jets.

This discussion is not an attempt at detailed modeling of the result
presented in the previous section; this will be done elsewhere.  
Neither shall we systematically invoke multi-wavelength observations.
Our purpose is rather to point out the multiple facets of
the \gr  phenomenon on its own.  They stem from the fact that the emission
probably originates from a population of accelerated particles
within a spatially confined relativistic jet, specifically oriented
towards the observer, and that subsequently the radiation must propagate
through the diffuse extragalactic background radiation field (DEBRA)
before it reaches us. Each of these circumstances can
influence the characteristics of the emitted spectrum, and we shall address
them in turn below.

\subsection{Production and absorption of TeV photons in the jet}

The enormous apparent VHE \gr luminosity of the source, reaching $\sim
10^{45} \, \rm erg/s$ during the strongest flares which typically last
$\Delta t = \, 1$ day or less, implies that the \grs are most probably
produced in a relativistic, small-scale (sub-parsec) jet which is directed
along the observer's line of sight. The determining quantity is the
so-called Doppler factor $\delta_{\rm j}= [(1+z) \Gamma_{\rm
j}(1-\beta_{\rm j}\cos{\Theta})]^{-1}$, where 
z is the redshift of a source
moving with velocity $\beta_{\rm j}c$ and Lorentz factor $\Gamma_{\rm
j}=(1-\beta_{\rm j}^2)^{-1/2}$ along the jet axis that makes an angle
$\Theta$ with the direction to the observer. Moreover, the assumption of
relativistic bulk motion appears to be unavoidable in order to overcome
the problem of severe $\gamma -\gamma$ absorption 
by pair production on low-frequency 
photons inside the source (see e.g. Dermer \& Schlickeiser \cite{Der:94}).
Indeed, assuming that the $\gamma$-radiation is emitted isotropically in
the frame of a relativistically moving source, the optical depth at the
{\it observed} \gr energy is easily estimated as
\begin{equation}
\tau_{\gamma \gamma}
\simeq \frac{f_{\rm r} d^2 \sigma_{\rm T} \delta_{\rm j}^{-6}\,E}{8m_{\rm
e}^2 c^6 \Delta t} \simeq 0.065 f_{-10} \Delta t_{\rm day}^{-1}
\delta_{10}^{-6} H_{60}^{-2} E_{\rm TeV}
\label{tau}
\end{equation}
Here $\delta_{10}=\delta_{\rm j}/10$, 
$\Delta t_{\rm day}=\Delta t /1 \,
\rm day$, $E_{\rm TeV} = E/1 \, {\rm TeV}$, with $E$ the energy of the 
gamma-ray in the laboratory frame, and 
$f_{-10} = f_{\rm r}/10^{-10}\, \rm erg/cm^2 s$ is the observed
energy flux at $h \nu \simeq 100 \delta_{10}^2 \, E_{\rm
TeV}^{-1} \, \rm eV$ which for an order of magnitude estimate is assumed to be
constant at the optical to UV wavelengths that predominantly contribute to the 
\gr absorption. Furthermore $d = cz/H_0 = 170 H_{60}^{-1}$ Mpc is the
distance to the source with redshift $z = 0.034$, 
normalized to the value of the Hubble constant
$H_{60} = H_0/60 \, {\rm km/s Mpc}$.
Assuming now that the observed optical/UV flux of Mkn 501,
$f_{-10} \simeq 0.5$ (see e.g. Pian et al.\ \cite{Pian:98}), 
is produced in the jet,
the absorption of 20 TeV \grs becomes negligible 
only when $\delta_{\rm j} \geq 10$; 
already for $\delta_{\rm j}=8$ internal absorption would be
catastrophic  for $\Delta t_{\rm day} = 1$, with $e^{-\tau_{\gamma \gamma}}
\leq 10^{-2}$.
One may interpret the detected VHE
spectrum of Mkn~501 given by Eq.~\ref{pexp} as a power-law {\it production} 
spectrum, modified by
internal $\gamma - \gamma$ extinction with optical depth 
$\tau = E_{\rm TeV}/6.2$.
This would give an accurate determination of the jet's Doppler factor
taking into account the very weak dependence of $\delta_{\rm j}$ on all
relevant parameters, namely $\delta_{\rm j}=8.5 \, f_{-10}^{1/6}
\Delta t_{\rm day}^{-1/6}\, H_{60}^{-1/3}$. Since there could be a number
of other reasons for the steepening of the TeV spectrum,
that estimate can only be considered as a robust lower limit on
$\delta_{\rm j}$.

The strong steepening of the observed spectrum of 
Mkn~501 above several TeV could also be attributed, for example, to an 
exponential cutoff in the spectrum of accelerated
particles. These could either be protons producing \grs 
through inelastic
$p-p$ interactions and subsequent $\pi^0$-decay ,  or electrons producing
\grs via
inverse Compton (IC) scattering. In IC  models 
an additional steepening of the \gr spectrum is naturally expected
due to the production of TeV \grs in the Klein-Nishina regime if the
energy losses of electrons are dominated by synchrotron radiation
in the jet's magnetic field. 
And finally, the exponential cutoff in the observed TeV spectrum
could be caused by $\gamma$-$\gamma$ absorption of 
TeV photons  in a possible
dust torus surrounding the AGN (Protheroe \& Biermann \cite{PrBr:96})
and in the extragalactic diffuse background radiation 
(Nikishov \cite{Nik:62}, Gould \& Schreder \cite{Gould:65}, 
Jelly \cite{Jelly:65}, Stecker et al. \cite{Steck:92}).  

\vspace{2mm}

Our current poor knowledge about the distortion of the source spectrum
caused by internal and intergalactic absorption does not 
allow us to distinguish
between hadronic and leptonic source models on the basis of their
predictions concerning the TeV energy spectra. Fortunately the temporal
characteristics are to a large extent free from these uncertainties. Thus
we believe that real progress in this area can only be achieved by the
analysis of both the {\it spectral and temporal} characteristics of X-ray
and TeV \gr emissions obtained during {\it multiwavelength} 
campaigns that investigate several X-ray selected BL Lac 
objects in different states of
activity, and located at different distances within several 100 Mpc.
Notwithstanding this belief, we  show here that the high-quality
HEGRA spectrum of Mkn~501 {\it alone} allows us to make quite a few
interesting inferences about the \gr production and absorption mechanisms.

\subsubsection{IC models of the gamma ray emission.}
Currently  it is believed (e.g. Ulrich et al. \cite{umu:97}) 
that the correlated 
X-ray/TeV flares  discovered by multiwavelength observations 
of Mkn~421 (Takahashi et al. \cite{tak:96}, Buckley et al. \cite{buk:96}) 
and Mkn~501 (Catanese et al. \cite{cat:97}, 
Pian et al. \cite{Pian:98}, Paper 1),
support the hypothesis  of both emission components
originating in  relativistic jets due 
to synchrotron/IC radiation of the 
same population of directly accelerated  electrons
(Ghisellini et al.\ \cite{GhisMD:1996}; Bloom \& Marscher \cite{BlMar:96};
Inoue \& Takahara \cite{IT:1996};
Mastichiadis \& Kirk \cite{Mast:97};
Bednarek \&  Protheroe \cite{BedPro:97}).   One of the distinctive
features of leptonic models is that they  allow 
significant temporal and spectral variations of TeV radiation.
The stable shape of the TeV spectrum of Mkn~501 
during the 1997 outburst does not
contradict these models. It rather requires them to have two important 
features.

First of all the form of the spectrum of accelerated electrons should be
essentially stable in time and be independent of the strength of the flare
up to electron energies responsible for the production of the highest
observed \gr energies $E \geq  16 {\rm TeV}$. For  a typical Doppler factor
$\delta_{\rm j} \sim 10$ this implies relatively modest electron energies
$E_{\rm e} \leq 10 \, {\rm TeV}$ in the jet frame, assuming that the Compton
scattering at the highest energies takes place in the Klein-Nishina limit.
In this limit a significant fraction of the electron energy goes to the
upscattered photon, i. e. $E \sim \delta_{\rm j} \, E_{\rm e}$.

For low values of the magnetic field in the emitting plasma, i.e. 
B { \raisebox{-0.5ex} {\mbox{$\stackrel{<}{\scriptstyle \sim}$}}} 
0.01~G, the X-ray spectrum could be more sensitive to accelerated 
electrons with energies above 10 TeV.
The typical observed energy of X-rays produced by
electrons of energy $E_{\rm e}$ in the jet frame is
\begin{equation}
E_{\rm X} \simeq 20 (B/0.1 \, {\rm G}) \, (E_{\rm e}/1 \, {\rm TeV})^2 \, 
\delta_{10} \, {\rm keV} .
\label{Ex}
\end{equation}
The BeppoSAX observations of Mkn~501 in April 1997 showed that 
 the X-ray spectrum becomes very hard during strong flares.
This is
interpreted as a shift of the synchrotron peak to energies in excess of
100 keV (Pian et al. \cite{Pian:98}). Formally this effect could be 
explained by 
a significant increase in each of the three parameters which determine the
position of the synchrotron peak, i.e. the maximum electron energy
$E_{{\rm e}, max}$, the magnetic field $B$, and the jet Doppler factor
$\delta_{\rm j}$.

The rather stable energy spectrum of TeV radiation
implies that the spectrum of the parent electrons
does not significantly vary during the HEGRA observations, the latter
performed with typical integration times between one and two hours.
The condition of a constant acceleration spectrum does not yet
guarantee  a stable energy spectrum of TeV radiation.
Therefore we need a second condition, namely to assume
very effective radiative (synchrotron and IC) cooling of
electrons, sufficiently fast  to establish
an equilibrium electron spectrum within
$\Delta t^{\ast}=10 \, \delta_{10} \, \rm h$. The radiative cooling time is
\begin{equation}
t_{\rm rad}= (\frac{4}{3} \, \sigma_{\rm T} \,
c \, w_0 \, E_{\rm e}/m_{\rm e} c^2)^{-1} \simeq 
\label{time}
\end{equation}
\[
\hspace*{0.8cm}
\simeq
15.5 \, (w_0/1 \, {\rm erg/cm^3)^{-1}}
\, E_{\rm TeV} \, \rm s,
\]
where $w_0=B^2/8 \pi + w_{\rm r}$ is the  total energy density of
magnetic and photon fields. Thus, for a  jet magnetic field of about
$0.1 \, \rm G$ and a comparable low-frequency photon density
($\approx 4 \cdot 10^{-4} \, \rm erg/cm^3$) a  radiative
cooling time of less than 5 hours (in the jet frame) could be easily
achieved.

\subsubsection{$\pi^0$ origin of gamma rays}
The lack of correlation between spectral shape and absolute flux,
as well as the very fact that \grs with energy $\geq 16 \, \rm  TeV$ 
are observed, could also be
explained, perhaps even in a more natural way, by the assumption of a
`$\pi^0$-decay' origin of the $\gamma$ -radiation. Yet the efficiency of
this
mechanism in the jet appears to be too low to explain the
observed time variability and the high fluxes of the TeV radiation. 
This is due to the low density $n_{\rm H}$ of the thermal 
electron-proton plasma in the jet. The problem of variability  
could be at least in
principle overcome by invoking adiabatic losses caused by 
relativistic
expansion of the emitting ``blob''. However, this assumption implies very
inefficient \gr production with a luminosity $L_{\gamma}=
L_{p}t_{\rm ad}/t_{\rm pp}^{\pi^0}$, where $L_{\rm p}$ is the luminosity in
relativistic protons, $t_{\rm ad}\geq R/c \sim \Delta t \delta{\rm j}$ is the
adiabatic cooling time, and $t_{\rm pp}^{\pi^0} 
\sim 5 \cdot 10^{15}(n_{\rm H}/1
{\rm cm}^{-3})^{-1} \, s$ denotes the characteristic
emission time scale of $\pi^0$-decay 
$\gamma$-rays. We shall assume here that the
proton luminosity $L_{\rm p}$
should not exceed the total power of the central engine, roughly
the Eddington luminosity $L_{\rm E} = 1.3 \cdot 10^{45} \, (M/M_{\odot})
\, \rm  erg/s$ of a
supermassive Black Hole of mass
$M=10^7 M_{\odot}$, or more empirically, the apparent ($4 \pi$) total
luminosity of the source
which is $L_{\rm tot} = 4\cdot {\pi}d^2\cdot f_{\rm tot} \sim 10^{45}$ erg/s,
where $f_{\rm tot} \sim {\rm few} \times 
10^{-10} \, \rm erg /cm^2 s$ is the total observed radiative flux  
(see e.g. Pian et al. \cite{Pian:98}).
Then the observed
TeV-flux of about $2.5 \cdot 10^{-10}$ erg/s requires a lower limit $n_{\rm H}
\geq 10^6 {\rm cm}^{-3}$ on the
density of the thermal plasma in the jet. This makes the
relativistically moving `blob' very heavy ($M \sim 0.05 M_{\odot}$) with
an unacceptably large kinetic energy $E_{\rm kin} = M c^2 \Gamma_{\rm j} 
\simeq 10^{54} \Delta t_{\rm day}^3 \delta_{10}^4 (n/10^{6} \, \rm cm^{-3})$ 
erg.

We would like to emphasize that these arguments hold 
against the $\pi^0$-origin
of \grs produced in a small-scale jet; they do not in general exclude
hadronic models. In particular, scenarios like the one assuming
$\gamma$-radiation produced by gas clouds that move across the jet 
(Bednarek \&  Protheroe \cite{BedPro:97}),
Dar \& Laor \cite{dar:97}) remain an attractive possibility for
hadronic models. They do not exclude either a ``proton blazar''
model  (Mannheim \cite{Man:93}).  It implies a secondary origin 
of the relativistic electrons that are the result of electromagnetic 
cascade, triggered by photo-meson processes involving extremely high 
energy protons in a hadronic jet (see Mannheim \cite{Man:98} and references
therein).

\subsection{Intergalactic extinction}
 
The effect of intergalactic  extinction of VHE 
\grs in diffuse extragalactic background radiation fields 
became  astrophysically
significant 
(see e.g. Stanev \& Franceschini \cite{Stanev:98},
Funk et al. \cite{Magn:97},
Biller et al. \cite{Bill:98},  
Stecker \&  de Jager \cite{Stec:98}, Biller \cite{vbill:98},
Primack et al.\ \cite{vprim:98}, Stecker \cite{vsteck:98}) 
after the discovery of TeV radiation from Mkn 421 and Mkn 501 up
to energies of 10 TeV, as
reported by the Whipple 
(Zweerink et al. \cite{Zv:97}), 
and HEGRA (Aharonian et al. 1997a), 
CAT (Djannati-Atai et al.\ \cite{CATcorrelation}),
and Telescope Array (Hayashida et al. \cite{hay:98}) groups. 

\subsubsection{Gamma ray absorption}

If we ignore the appearance of second generation $\gamma$-rays 
(see Sect. 6.2.2),
then extinction is reduced to a simple absorption effect which can 
be described by a single absorption optical depth $\tau$.  

The optical depth $\tau$ of the intergalactic medium for a 
$\gamma$-ray photon of energy $E$, emitted from a source at the distance 
$d=c z/H_0$, can be  expressed 
for small redshifts z$\ll$ 1
in a convenient approximate
form using a quantity $\tau=\tau^{\prime} \xi$, where
\begin{equation}
\tau^{\prime}(E) = \frac{\sigma_{\rm T}}{4} \,      
\epsilon_{\rm m} n(\epsilon_{\rm m}) \, d  \simeq
\label{taup}
\end{equation}
\[
\hspace*{5mm}
0.08 \left(\frac{\epsilon_{\rm m}^2 n(\epsilon_{\rm m})}{10^{-3} \, 
\rm eV/cm^3)}\right) \, 
\left(\frac{z}{0.034}\right) 
E_{\rm TeV}  \, H_{\rm 60}^{-1}
\]

\noindent
with
$H_{\rm 60}=H_0/60 \, \rm km/s \, Mpc$,
$E_{\rm TeV}=E/1 \, \rm TeV$,
$\epsilon_{\rm m}=4 m^2_{\rm e} c^4 /E \simeq 1 E_{\rm TeV}^{-1} \, 
\rm eV$,  and
$\xi$ being a correction factor 
which accounts for the specific form of the
differential DEBRA photon number density $n(\epsilon)$; the
background
photon energy is denoted by $\epsilon$.
This expression is based on the narrowness  
of the $\gamma \gamma \rightarrow e^+ e^-$ cross-section $\sigma_{\gamma
\gamma}$ as a function of $(\epsilon /\epsilon_{\rm m})$ which peaks at
$\epsilon / \epsilon_{\rm m} \simeq 1$  in an
isotropic 
field of background photons (Herterich \cite{her:74}). Thus for a large
class of 
broad DEBRA spectra $n(\epsilon)$ the optical depth is essentially
caused 
by background photons with energy centered around $\epsilon_{\rm m}$.
For (broad) power-law spectra, 
$n(\epsilon)=n_0  \epsilon^{-\gamma}$, 
the optical depth can be calculated analytically as
$\tau(E)=\eta(\gamma) \cdot 4^\gamma (\sigma_{\rm T}/4) 
\epsilon_{\rm m} n(\epsilon_{\rm m}) \, d$, where  
$\eta(\gamma)=7/6 \gamma^{-5/3} (1+\gamma)^{-1}$
(Svensson \cite{sven:87}).  
Thus for relatively flat power-law spectra with $1 < \gamma < 2.5 $
we obtain 
$\xi=4^\gamma \eta(\gamma) \simeq 2$, i.e. approximately half 
of $\tau_{\gamma \gamma}$ is contributed by background 
photons 
with $\epsilon$ in the interval given by $\epsilon_{\rm m} \pm 1/2
\epsilon_{\rm m}$.

\begin{figure}
\resizebox{8.1cm}{!}{\includegraphics{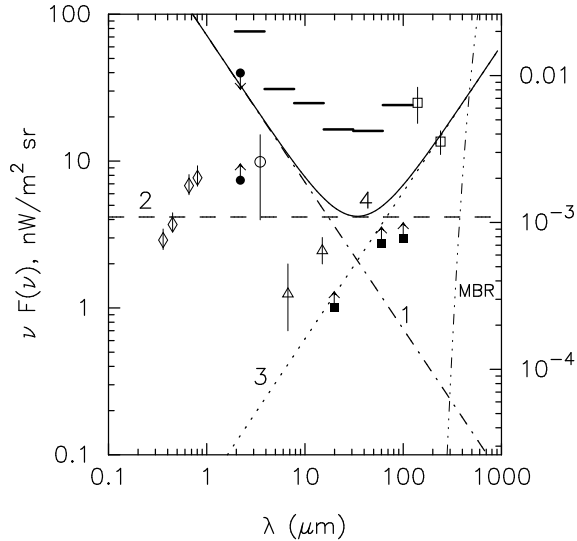}}
\caption
{The energy fluxes of DEBRA for pure  power-law differential spectra 
with $\gamma=1$ (curve 1), $\gamma=2$ (curve 2) and 
$\gamma=3$ (curve 3). Curve 4 is the sum of spectra 1 and 3.
The absolute flux normalizations have been determined from 
the condition of $\tau_{\gamma \gamma}=3$ (for curve 1), and
from the condition of the maximum possible flux of DEBRA which still
``reproduce'' reasonable \gr source spectra shown in Fig.~ \ref{fig2_dis}
(for curves 2 and 3). The horizontal bars correspond to the 
upper limits on DEBRA fluxes obtained using a method similar 
to the one suggested  by Biller et al. (1998).
The curve marked as ``MBR'' correspond to the density of the
2.7 K MBR. 
The tentative flux measurement at 3.5 $\mu$m is taken from Dwek \& Arendt
(1998), and the flux estimates based on the ISO survey at 6 and 15 $\mu$m
are from Stanev \& Franceschini (1998). The other measured fluxes and the
upper/lower limit estimates of the DEBRA are taken from the recent
compilation by Dwek et al.\ (1998), i.e., the upper limit at 2.2 $\mu$m is from
Hauser et al.\ (1998), the lower limit at 2.2 $\mu$m is from Gardner 
et al.\ (1997), and the UV to optical detections are from Pozzetti et al.
(\protect\cite{Poz:98}).}   
\label{fig1_dis}
\end{figure}

We note that $\tau(E) \propto E^{\gamma
-1}$ for a power-law spectrum of DEBRA .
For example, within the `valley' of the 
energy density at mid infrared 
wavelengths from  several $\mu \rm m$ to several tens of $\mu \rm m$,
where the
energy density is expected to be more or less constant (i.e. $\epsilon^2
n(\epsilon)= {\rm const}$),
the intergalactic extinction of \grs is largest
at the highest observed energies. In particular, according to Eq.~\ref{taup},  
even at a very low and probably unrealistic level of the DEBRA intensity
of  
$\epsilon^2 n(\epsilon)=10^{-4} \, \rm eV/cm^3$  at $\lambda \sim 30 \,
\rm \mu m$, 
approximately $35 \, \rm percent$ of the 25 TeV $\gamma$-rays emitted by 
Mkn~501 are extinguished before they reach the observer.
This implies that the observed TeV spectrum of Mkn~501 
contains important information about the DEBRA,
at least at wavelengths  $\lambda \geq 10 \mu \rm m$. 
Moreover, it is quite possible that a non-negligible intergalactic
extinction  takes place also at {\it low} \gr energies due to 
interactions with  near infrared (NIR) background photons. 
Interpreting 
the power-law shape of the spectrum at $2 \, \rm TeV$ as an 
indication for weak
extinction by, say, a factor less than or equal to 2, and ignoring
the contributions
from all other wavelengths beyond the interval $0.5 \pm 0.25 \, \rm eV$
($\lambda \simeq 2.5^{\rm +2.5}_{\rm -0.85} \,  \, \mu \rm m$) 
from Eq. \ref{taup}, one obtains $\epsilon^2 n(\epsilon) \leq 4 
\times 10^{-3} \, H_{\rm 60} \; \rm eV/cm^3$,
not far from the flux of DEBRA experimentally inferred  
(Dwek et al. \cite{dwek:98},  de Jager \& Dwek \cite{gt:98})
and theoretically expected (Malkan \& Stecker \cite{malk:98},
Primack et al. \cite{vprim:98}) at these  wavelengths.

However, to the extent that the source spectrum of the \grs is unknown, 
this cannot be considered as a model-independent upper limit
(e.g. Weekes et al. \cite{GRO4rev:97}). 
Obviously for any quantitative estimate
of the DEBRA one needs to know the intrinsic 
\gr spectrum, and this can not be obtained from \gr 
observations alone. As already emphasized
above, a multi-wavelength approach is
indispensable, for example in the form of detailed modeling of the
{\it entire} or at least a large wave-length range of the nonthermal
spectrum. To be specific, one avenue would be modeling the nonthermal 
X-ray and \gr spectra in the framework of synchrotron-inverse
Compton models, based on simultaneous multi-wavelength observations of
X-ray selected BL Lac objects (Coppi \& Aharonian \cite{CoAh:98}).

\begin{figure}
\hspace*{0.5cm}
\resizebox{7.3cm}{!}{\includegraphics{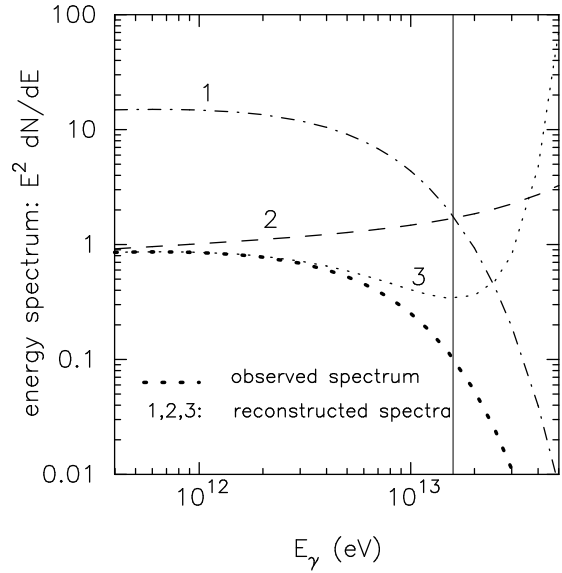}}
\caption
{The source spectra of Mkn 501 reconstructed for 
different models of DEBRA. The heavy dots correspond to the
measured spectrum of HEGRA approximated by Eq.~6 with
$\gamma=1.92$, and $E_0=6.2 \, \rm TeV$. The curves
1, 2, and 3 correspond to the power-law DEBRA spectra
shown in Fig. \ref{fig1_dis} by curves 1,2 and 3, respectively. 
The vertical line at 16 TeV indicates the edge of the
\gr spectrum of Mkn 501 measured by HEGRA.}
\label{fig2_dis}
\end{figure}

An illustration for the need of more than \gr data alone is given by a 
DEBRA spectrum $n(\epsilon) \propto \epsilon^{-1}$.
It does not change the spectral shape of \grs at all (grey opacity),
although formally
the extinction could be arbitrarily large. The curves marked as ``1''
in Fig.~\ref{fig1_dis} and Fig.~\ref{fig2_dis} may serve as an example.
In the case of Mkn~501 this ambiguity can 
be significantly reduced by rather general
arguments regarding
$\gamma$-ray {\it energetics}. Requiring again
that the \gr luminosity  should not exceed the total apparent
luminosity of the  source, $L_{tot} \sim  10^{45} \, \rm erg/s$,
we must have $\rm e^{\tau} \leq 10^4 (\delta_{\rm j}/10)^{4}$.
For \gr production in the jet with $\delta_{\rm j} \sim 10$,
this implies $\tau \leq 10$. 
In fact,
already a value of $\tau \sim 3$
(corresponding to curve 1 of Fig.~\ref{fig1_dis}) 
creates uncomfortable conditions for
the majority of realistic
models of high energy radiation from Mrk~501, assuming that the
nonthermal emission is produced in the jet due to 
synchrotron and inverse Compton processes. 
Indeed,  $\tau_{\gamma \gamma} \sim 3$ implies that 
the $\gamma$-ray luminosity of the source corresponding
to the ``reconstructed spectrum'' (curve 1 in Fig.~\ref{fig2_dis})
exceeds  the luminosity of the source in all other wavelengths 
by an order of magnitude
which hardly could be
accepted for any realistic combination of parameters
characterizing the jet.

Accepting the current lack of reliable knowledge of the \gr source
spectrum, it is nevertheless worthwhile 
to derive  upper limits on DEBRA by formulating 
different {\it a priori}, but {\it astrophysically} meaningful 
requirements on the spectrum and the \gr luminosity of the source.
A possible criterion, for example,  could be that   
within any reasonable source model the intrinsic spectrum of 
$\gamma$-rays, $J_0(E)=J_{\rm obs} \, \exp(\tau)$, 
should not contain a feature which exponentially rises with energy 
at any {\it observed} \gr energy. In practice this
implies
that the `reconstruction' factor  $\exp(\tau)$ should not significantly 
exceed  the exponential term
of the observed \gr spectrum from Eq.~6. This condition is most directly 
fulfilled by the power  
law-spectrum with $\gamma = 2$ that has equal DEBRA power per unit
logarithmic bandwidth in energy. It results in $\tau (E)\propto E$ and
then yields an upper limit for the DEBRA density close to   
$\epsilon^2 n(\epsilon)=10^{-3} \rm eV/cm^3$. This limit corresponds
to the  borderline on which the source spectrum becomes a pure
power-law  Figs. \ref{fig1_dis} and \ref{fig2_dis}, curves 2. 
A slight increase of the DEBRA density by as little as a factor
of 1.5 leads to a dramatic (exponential) deviation 
of the reconstructed spectrum at the highest observed \gr 
energies around 16 TeV from the  $E^{-1.9}$ power-law extrapolation.

A power law with $\gamma = 3$
would give similar and complementary results. The resulting source spectra
(Fig.~\ref{fig2_dis}) and  the upper limits on the
DEBRA density (Fig.~\ref{fig1_dis})
obtained in this way assuming power-law
spectra for DEBRA with
$\gamma = 2$ and 3,  are given by the curves 2 and 3, respectively. The power
law $\gamma = 1$ corresponding to $\tau_{\gamma \gamma}=3$ 
complements Fig.~\ref{fig1_dis}.

\begin{figure}
\resizebox{8.1cm}{!}{\includegraphics{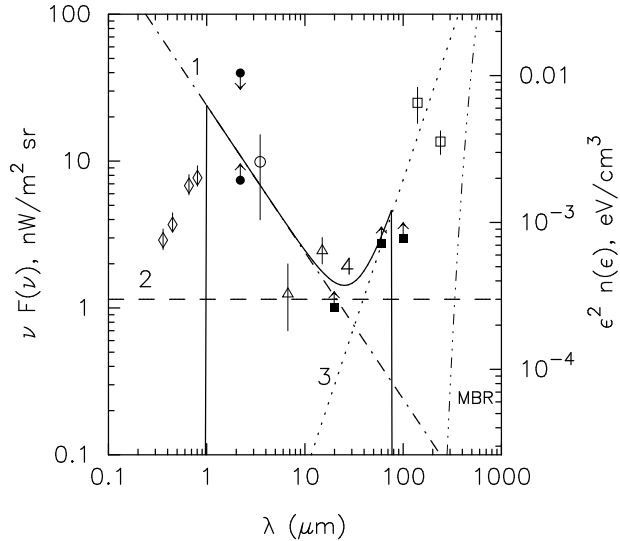}}
\caption
{The same as in Fig.~\ref{fig1_dis}, but for different parameters of the
power-law DEBRA. Curve 1: $\gamma=1$ with an absolute flux
corresponding to $\tau_{\gamma \gamma}=1$; curve 2:
$\gamma=1$ with $\epsilon^2 n(\epsilon)=3 \cdot 10^{-4} \, \rm eV/cm^3$;
curve 3: $\gamma=4$; curve 4: sum of spectra 1 and 3,
truncated at 1 and 80 $\mu$m.}
\label{fig3_dis}
\end{figure}
It should be noted however that any realistically expected spectrum 
of the DEBRA  in a broad range of wavelengths
deviates from a simple power-law. In fact, all
models of the DEBRA, independently of the details, predict two pronounced
peaks
in the spectrum at 1~$\mu $m and 100~$\mu $m 
contributed by the emission of the stars and of the interstellar dust,
respectively,
and a relatively flat `valley' at mid IR wavelengths around 10~$\mu$m 
(see e.g. Dwek et al. \cite{dwek:98}). The strong impact of the 
DEBRA spectrum on calculations of the opacity of the intergalactic 
medium has been emphasized  by Dwek \& Slavin (\cite{DwSl:1994}) and
Macminn and Primack (\cite{Pri:96}).

\begin{figure}
\hspace*{0.5cm}
\resizebox{7.3cm}{!}{\includegraphics{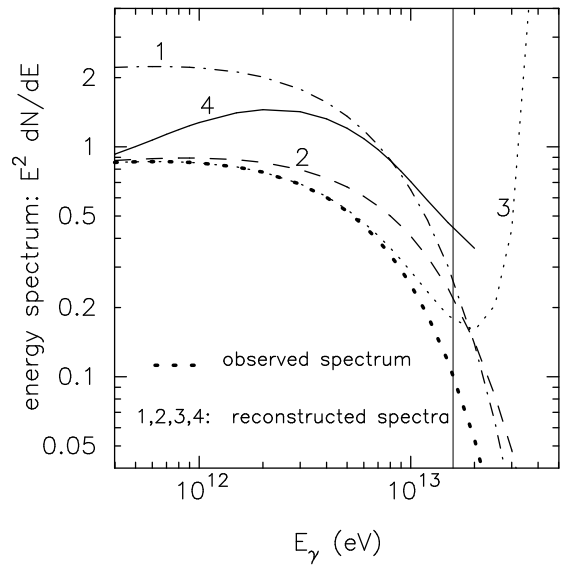}}
\caption
{The same as Fig.~\ref{fig2_dis} but for DEBRA fluxes 
shown in Fig.~\ref{fig3_dis},  with the addition of a reconstructed 
spectrum (curve 4) corresponding to curve 4 in Fig. \ref{fig3_dis}.
}
\label{fig4_dis}
\end{figure}

Note that the power law with $\gamma \sim 1$
characterizes the shape of the spectrum of DEBRA at 
near IR wavelengths, typically between 1 and several microns,
and the power-law with $\gamma \geq 3$ characterizes the 
DEBRA between 10 and 100 microns. Interestingly,  the sum of these 
two power law spectra  ``reproduces'' a reasonable shape of the `valley'. 
This is seen  in Fig.~\ref{fig1_dis} 
and Fig.~\ref{fig3_dis} where the measured fluxes or estimated 
upper and lower limits  obtained directly at different wavelengths of 
the DEBRA are shown.

Finally, an interesting numerical criterion for the derivation of upper
limits on the DEBRA was suggested by Biller et al. (\cite{Bill:98}). 
It relies  on independent $\gamma$-ray observations. 
A variant of this approach, where the
additional restriction forbids a  differential source spectrum
harder than $\propto E^{-1.5}$  within the observed energy range, is
shown by the horizontal bars in Fig.~\ref{fig1_dis}.

The results described above could be ``improved'' assuming a more realistic,
$\tau_{\gamma \gamma}=1$ optical depth for the $\gamma = 1$ power law 
branch at short wavelengths, and a steeper, $\gamma = 4$, power 
law branch at long
wavelengths (see Fig.~\ref{fig3_dis}). 
The latter choice for $\gamma$ is due to 
the rapid rise of the data points
towards far infrared wavelengths $\lambda$ in order to
fit the recent measurements of the flux at $140 \, \mu \rm m$
by DIRBE aboard the COBE satellite. 
For the far infrared (FIR) branch
the absolute flux of the power-law  with $\gamma = 4$
is chosen again from the condition 
that the differential $\gamma$-ray source spectrum should not exponentially
rise at energies up to $E \sim 16 \, \rm TeV$ 
(see Fig.~\ref{fig4_dis}).  Note also that the criterion 
of $\tau_{\gamma \gamma}=1$
for the $\gamma = 1$ branch at short wavelengths
is pretty close to the level of the flux of the recent tentative 
detection of DEBRA at $3.5 \, \mu \rm m$ (Dwek \&  Arendt \cite{DwAr:98}). 
The sum of the NIR and FIR power-law branches with the above
indices and absolute fluxes results in a deeper mid IR
``valley'' and predicts a very steep spectrum of DEBRA
between 30 and 100 $\mu \rm m$.  In Figs.~\ref{fig3_dis} and \ref{fig4_dis}  this
spectrum has been  truncated at $\lambda \simeq 80 \mu \rm m$
that  corresponds to the kinematic 
threshold of pair production at interactions
with the maximum observed energy of \grs  of about 20 TeV.
The spectrum is also truncated at $\lambda=1 \mu \rm m$
in order to avoid significant excess compared 
with the fluxes at optical/UV wavelengths
recently derived from the  Hubble Deep Field analysis 
(Pozzetti et al. \cite{Poz:98}).

The effect of ``reconstruction'' of the \gr spectra of Mkn 501 
corresponding to this  ``best estimate''  of DEBRA 
between 1 and 80 $\mu$m  is illustrated
in Fig.~\ref{fig4_dis}. It shows, in the simple absorption 
picture  (using two truncated power-laws) that even a
conservative choice for the DEBRA field implies intergalactic extinction
at all observed energies.  Especially  the reconstructed spectrum around
1 TeV could be considerably harder than the observed spectrum, with a
maximum of $E^2 \, {\rm d}N/{\rm d}E$ at  2 TeV
(see Fig.~\ref{fig_spectrum}). This also
demonstrates that it would be dangerous to interpret the observed spectral
slope at low energies in terms of a power law extending from still lower
energies.

\subsubsection{The effect of cascading in the DEBRA}
The discussion of intergalactic  $\gamma - \gamma$ absorption effects is
in principle incomplete without considering secondary radiations. Briefly,
when a \gr is
absorbed by pair production, its energy is not lost. The secondary
electron-positron pairs create new \grs via inverse Compton scattering on
the 2.7 K MBR. The new \grs produce more pairs, and thus an
electromagnetic cascade develops (Berezinsky et al. \cite{Berez:90}, 
Protheroe \& Stanev \cite{Prostan:93},
Aharonian et al.\ \cite{AhCo:94}). 
In fact, in our discussion
of absorption  we have neglected any secondary \grs in the field
of view of the detector and we finally turn now to
these.

For a  primary \gr
spectrum ${\rm d}N/{\rm d}E$ harder than
$E^{-2}$, extending to energies $E \gg 1 {\rm TeV}$, the cascade spectrum
at TeV energies
could strongly dominate over the primary \gr spectrum.
In addition, the spectrum of
the cascade \grs that reach the observer 
has a standard form independent of the 
primary source spectrum with a characteristic 
photon index of $1.8-2.0$ at energies between $\sim 100~\rm GeV$
and an exponential cutoff determined by the 
condition $\tau(d,E)=1$.
Thus, somewhat surprisingly, the  measured time-averaged spectrum
of Mkn~501 can in principle be fitted
by a cascade \gr spectrum for a reasonable DEBRA flux level of about
$10^{-3} \, \rm eV/cm^3$, provided that 
the {\it invisible} source spectrum extends 
well beyond 25 TeV, say up to 100 TeV.
 
For an intergalactic magnetic field (IGMF) $B > 10^{-12} \,
\rm G$
the cascade \grs could be observed in the form of an extended emission
from a giant pair halo with a radius up to several degrees formed around
the
central source (Aharonian et al.\ \cite{AhCo:94}). 
Although the possible
extinction of \grs from Mkn~501, at least above 10 TeV, unavoidably
implies the formation of a pair halo, the TeV radiation of Mkn~501 cannot
be attributed to such a halo simply due to arguments based on the
detected angular size
and the time variability of the radiation.

However, the speculative assumption of an extremely low IGMF
still allows an interpretation of the observed TeV \grs of
Mkn~501 within the hypothesis of a cascade origin. 
Instead of extended and persistent halo radiation, 
we expect in this case that the cascade \grs penetrate from the source to the
observer almost on a straight line.  Yet at cosmological
distances to the source even very
small deflections of the cascade electrons by the IGMF lead to significant
time delays of the arriving $\gamma$-rays:  
$\Delta t_{\rm B} \simeq 2.4 (d/1 \,
{\rm
Gpc})(E/1 \, {\rm TeV})^{-2} (B/10^{-18} \, {\rm G})^2 \, \rm days$ 
(Plaga \cite{Plaga:95}).  
This implies that in order to see more or less synchronous
activity (within
several days or less)  of Mkn~501 at different wavelengths, as
was
observed during multiwavelength observations of the source (see e.g. Pian
et al. \cite{Pian:98}), we would have to require $B \leq 10^{-18} \, \rm G$.
Although indeed
quite speculative, such small magnetic fields on spatial  scales large
compared to
1 Mpc cannot be
{\it a priori} excluded (e.g. Kronberg \cite{Kron:96}). 
\section{Conclusions}
In this paper we have presented the 1997 Mkn 501 time averaged spectrum as
measured with the HEGRA IACT system in the energy range from 500 GeV to 24
TeV. The absence of strong temporal evolution as well as a
significant correlation of the emission strength and the spectral shape
in the energy region from 500 GeV to 15 TeV made the determination of a 
time averaged spectrum astrophysically meaningful. 
Due to unprecedented \gr statistics and the 20\% 
energy resolution of the instrument, it was possible to detect for the
first time $\gamma$-rays from an extragalactic source with 
energies well beyond 10~TeV, and to measure a smooth, curved energy 
spectrum deeply into the exponential regime. 
We found, that the spectrum above 0.5 TeV is well described by a power law
with an exponential cutoff $dN/dE$ = 
$\, N_0 \,E^{-1.92}\,\exp{(-E/6.2 \, \rm TeV)}$, 
the highest recorded photon energies being 16 TeV or more.

The detection of TeV $\gamma$-rays from Mkn~501 leads to the
unavoidable conclusion that the observed $\gamma$-radiation 
is produced in a  relativistic jet with a Doppler factor
$\delta_{\rm j} \geq 10$. Actually, assuming a pure power-law 
{\rm production} spectrum of $\gamma$-rays we may naturally explain
the exponential cutoff in the observed spectrum  spectrum 
by an internal $\gamma$-$\gamma$
absorption in the jet. Because of the strong dependence
of the optical depth on the jet's Doppler factor,
this hypothesis  gives an accurate determination
of the latter, $\delta_{\rm j}=8.5$. Remarkably, the 
uncertainty of this
estimate, which is mainly  due to the uncertainty 
in the value of the Hubble constant, $H_0 \simeq 60^{+40}_{-20} \, \rm
km/c Mpc$ and in the measured energy of the  
exponential cutoff $E_0=6.2 \, \rm TeV$,
does not exceed $20 \%$. However, since there could be other
reasons for the steepening of the TeV spectrum, this
estimate can only be considered as a robust lower limit on $\delta_{\rm j}$.
In particular, the steepening of the $\gamma$-ray spectrum at the highest
energies could be attributed to an exponential cutoff in the
spectrum of accelerated particles, as well as -- in the
case of the inverse Compton origin of $\gamma$-rays --
to the Klein-Nishina effect. 

In addition, a  modification of the 
intrinsic (source) spectrum of TeV 
$\gamma$-rays takes place during
their passage through the intergalactic medium. The recent 
claims about  tentative detections  of the diffuse 
extragalactic background radiation by 
the DIRBE instrument aboard the COBE at near infrared 
($\lambda=3.5 \mu \rm m$) and far infrared 
($\lambda=140 \mu \rm m$) wavelengths, both
at the $\simeq 10 \, \rm nW/cm^2 sr$ flux level
imply a strong effect of intergalactic $\gamma$-$\gamma$ extinction
on the observed Mkn~501 spectrum over the entire energy region 
measured by HEGRA. In particular,  the shape of the highest energy part 
of the observed $\gamma$-ray spectrum combined 
with the DIRBE flux at $140 \mu \rm m$ requires a very 
steep ($\nu F_\nu \propto \lambda^{s}$ with $\nu \sim 4$) spectrum of
the DEBRA with a characteristic flux at mid infrared wavelengths
($\lambda \sim 10-30 \mu \rm m$) around 1 to 2 $\rm nW/m^2 ster$.
Due to the expected  flat spectrum of DEBRA at near infrared
wavelengths (close to $\nu F_\nu \propto \lambda^{-1}$) the
modification of the $\gamma$-ray spectrum 
is less prominent  at energies of a few
TeV, although the absolute extinction could be
very large. This does not allow us to draw definite  conclusions
about the absorption effect based on the analysis of the
shape of the $\gamma$-ray spectrum. Nevertheless, in the 
case of Mkn~501 this ambiguity can 
be significantly reduced by rather general
arguments regarding the $\gamma$-ray {\it energetics}.
Indeed, even relatively modest assumption about 
the optical depth of about $\tau \sim 3$ which corresponds to
a flux of the DEBRA in the K-band ($\lambda=2.2 \mu \rm m$)
of about $30 \rm \, nW/m^2 ster$ creates uncomfortable conditions for
any realistic models of the high energy radiation from Mrk~501.
Thus, this value of the flux of the DEBRA may be considered as a
rather strong upper limit comparable with the DIRBE upper limit
at this wavelength.

To summarize, our excursion through the 
nonthermal physics of AGNs shows that \gr
observations alone do not allow a unique interpretation of a spectrum like
the one we have presented for Mkn~501, even though one can make a number
of highly interesting inferences. In particular, our current poor knowledge 
about the intrinsic spectrum of Mkn~501
does not allow us to make definite conclusions about the
effect of the intergalactic absorption of TeV $\gamma$-rays.
And {\it vice versa}, the uncertainty in the density of DEBRA
does not allow us to take into account the effect of the 
intergalactic absorption, and thus to ``reconstruct'' 
the intrinsic $\gamma$-ray spectrum.
The knowledge of the latter is an obvious condition
for the quantitative study of the acceleration and radiation
processes in the source. Therefore we believe that  decisive progress
in this field could
be achieved by the analysis of both the spectral and
temporal characteristics of X-ray and TeV $\gamma$-ray emissions
obtained during the multiwavelength campaigns of  several
X-ray selected BL Lac objects at different states of activity, and
located at different distances within 1 Gpc.
Due to the time variability
of such sources this requires simultaneous observations together with
extensive theoretical modeling. Given these we expect to obtain
independent knowledge of the diffuse intergalactic radiation fields which
can be compared with direct measurements and models of galaxy formation. 
We hope that the successful multiwavelength campaigns of Mkn 421
and Mkn~501 in 1998 with participation of several X-ray satellites
and the HEGRA IACT system will provide highly interesting results
in this area.\\[2ex]
                                                          
\noindent
{\it Acknowledgments}. The support of the German ministry for Research and
technology BMBF and of the Spanish Research Council CYCIT is gratefully
acknowledged. We thank the Instituto de Astrophysica de Canarias
for the use of the site and for supplying excellent working conditions at
La Palma. We gratefully acknowledge the technical support staff of the
Heidelberg, Kiel, Munich, and Yerevan Institutes. 

\appendix
\section{Systematic errors on the shape of the spectrum}
To illustrate the effect of systematic errors for determination
of energies and effective areas in more detail, we consider here the
following model. 
For simplicity, we consider a detector consisting of a single 
Cherenkov telescope, which is located inside a uniformly illuminated
Cherenkov light pool of area $A_{\rm pool}$. 
To good approximation, the light yield $I$ 
(in photons per m$^2$) is proportional to the shower energy,
\begin{equation}
I = \alpha E
\end{equation}
neglecting logarithmic corrections. The response of the telescope is
now characterized by two quantities, namely the total intensity $Q$
detected in the image (usually the {\em size},
measured e.g. in units of ADC channels, or, 
with a conversion factor, in units of `photoelectrons'), and the
signal $V$ induced in the highest pixel(s), which is fed into the
trigger circuitry. Both $Q$ and $V$ should be proportional to $I$,
\begin{equation}
Q = \beta I \, ; \, V = \gamma I \, ,
\end{equation}
but they are influenced by rather different factors. While both 
$\beta$ and $\gamma$ include the mirror reflectivity, the PMT quantum
efficiency, and the PMT gain, 
$V$ is to a much higher degree sensitive to the point
spread function of the mirror, and to the shape of the signal generated
by the PMT \footnote{In HEGRA, like in most other IACTs, the ADCs
effectively measure
the integral charge delivered by the PMTs, regardless of the exact
time-dependence of the signal. In contrast, the trigger
circuitry is voltage-sensitive, and a signal of a given charge
may or may not trigger, depending on the exact waveform.}. 
The effective detection area is given by
\begin{equation}
A = P(V) A_{\rm pool}
\end{equation}
where $P$ is the trigger probability for a given value of $V$.
The fact
that triggering and energy determination are not based on 
identical quantities is the key origin of systematic errors, in 
particular in the threshold region.

In the analysis of data, values $\alpha_{\rm MC}$, $\beta_{\rm MC}$ and
$\gamma_{\rm  MC}$ are assumed for these constants, usually based on 
Monte Carlo simulations. The values may differ from the true
values due to imperfections in the parameterization of the
atmosphere ($\alpha$) or of the optics and electronics of the
telescope ($\beta$, $\gamma$). The reconstructed energy $\tilde E$
of a shower of true energy $E$ is then, in the absence of fluctuations,
\begin{equation}
\tilde E = \alpha_{\rm MC}^{-1} \beta_{\rm MC}^{-1} Q =
\alpha_{\rm MC}^{-1} \beta_{\rm MC}^{-1} \beta \alpha E \equiv f E~.
\end{equation}
Based on eqs.~\ref{eqa} and \ref{eqb} and using 
${\cal R}(E',E)=\delta(E'-f\,E)$
the rate of events with measured energy $\tilde E$ is then
\begin{equation}
r(\tilde E) = \phi(\tilde E / f) P(\alpha 
\gamma \tilde E /f) A_{\rm pool} / f~.
\end{equation}
In the analysis, the Monte Carlo simulated rate $r_{MC}$ is used to
evaluate the effective area
\begin{equation}
r_{MC}(\tilde E) = \phi_{\rm MC}(\tilde E) P(\alpha_{\rm MC} 
\gamma_{\rm MC} \tilde E) A_{\rm pool} \, ,
\end{equation}
resulting in a reconstructed flux
\begin{equation}
\phi_{rec}(\tilde E) = \phi_{MC}(\tilde E) { r(\tilde E) \over
r_{MC}(\tilde E)}
\end{equation}
\[ 
\hspace*{1.4cm} = {1 \over f} \phi(\tilde E / f) 
{P(\alpha \gamma \tilde E /f) \over P(\alpha_{MC} \gamma_{MC} \tilde E)}~~~.
\]
Incorrect constants used in the simulation may hence result in 1) a
factor $f$ modifying the energy scale, but not the shape of the spectra,
and 2) a change of the shape of the spectra in particular in the 
threshold region, where $P(V)$ varies steeply with $V$. The scale 
factor $f$ cannot be determined from IACT data alone; some external
reference is required. The second effect 
-- the distortion of the spectra -- disappears provided that 
$\alpha \gamma /f = \alpha_{MC} \gamma_{MC}$, i.e., 
$\gamma_{MC} / \beta_{MC} = \gamma / \beta$, assuming that the simulation
correctly accounts for the statistical fluctuations determining the shape
of $P(V)$. This condition can be checked internally within the 
data set. In  particular, one should compare the distribution in $Q$ for 
a given value of $V$ (or for a fixed range in $V$, $V > V_0$) in the data 
and in the simulation (see Sect.\ 4.3 and Fig.\ 5). 
If $\gamma / \beta$ differs between data and simulation,
the distribution of $Q$ in the threshold region will be different. 
This comparison also tests the simulation of the shape of $P(V)$.

If the thresholds in $Q$ are shifted by a factor $1+\epsilon$ in the data 
relative to the simulation, this implies that $\gamma_{MC} / \beta_{MC}$ 
is off by the same factor, resulting in the flux error given by Eq.~\ref{eqd}.

\end{document}